\documentclass[aps,prl,longbibliography,twocolumn,superscriptaddress,amsfont,graphicx,nofootinbib,preprintnumbers,floatfix,10pt]{revtex4-1}%
\UseRawInputEncoding
\usepackage{ifpdf}
\usepackage{amsmath}
\usepackage{bm}
\usepackage{rotating}
\usepackage[english]{babel}
\usepackage{amssymb}
\usepackage{braket}
\usepackage{setspace}
\usepackage{booktabs}
\usepackage{fontawesome}
\usepackage{afterpage}
\usepackage{booktabs} 
\allowdisplaybreaks[4]

\usepackage{xcolor}
\usepackage{graphicx}
\usepackage{placeins}
\usepackage{hyperref}
\hypersetup{hypertexnames=false}
\usepackage{enumerate}
\usepackage{lipsum}

\usepackage{slashed}

\usepackage{changes}

\usepackage{mathtools}
\usepackage{siunitx}

\newcommand{\bB}{\mathbf{B}}
\newcommand{\bk}{\mathbf{k}}
\newcommand{\bn}{\mathbf{n}}
\newcommand{\dd}{\mathrm{d}}
\newcommand{\wpl}{\omega_{\rm pl}}
\newcommand{\nelec}{n_{\rm e}}
\newcommand{\kL}{k_{\rm L}}
\newcommand{\kH}{k_{\rm H}}
\newcommand{\kmin}{k_{\rm min}}

\newcommand{\rev}[1]{#1}

\begin{document}

\title{Enhancing Graviton--Photon Conversion through Galactic Magnetic Turbulence}

\author{Hongliang Tian}
\email{hongliangtian@njnu.edu.cn}
\affiliation{Department of Physics and Institute of Theoretical Physics, Nanjing Normal University, Nanjing, 210023, China}

\author{Lei Wu}
\email{leiwu@njnu.edu.cn}
\affiliation{Department of Physics and Institute of Theoretical Physics, Nanjing Normal University, Nanjing, 210023, China}
\affiliation{Nanjing Key Laboratory of Particle Physics and Astrophysics, Nanjing, 210023, China}

\author{Yang Zhang}
\email{zhangyang2025@htu.edu.cn}
\affiliation{Center for Theoretical Physics, Henan Normal University, Xinxiang 453007, China}

\author{Bin Zhu}
\email{zhubin@mail.nankai.edu.cn}
\affiliation{School of Physics, Yantai University, Yantai 264005, China}

\begin{abstract}
Plasma dispersion suppresses coherent graviton-photon conversion over astrophysical distances. We show that Galactic magnetic turbulence enables phase matching by compensating the plasma-induced mismatch, allowing matched magnetic modes to contribute coherently. In the long path limit, the ensemble-averaged conversion probability grows linearly with propagation distance in the phase matched regime and is governed by the projected magnetic power spectrum. Applied to photon background observations, this mechanism yields characteristic strain limits that are stronger than previous Galactic conversion bounds in the inertial range phase-matching window, with improvements of up to two orders of magnitude.
\end{abstract}

\maketitle

{\it Introduction.}~~
Ground-based interferometers~\cite{Gertsenshtein:1962kfm,LIGOScientific:2016aoc} and pulsar timing arrays~\cite{NANOGrav:2023gor,Xu:2023wog,Reardon:2023gzh,EPTA:2023fyk,Miles:2024seg} probe gravitational waves in widely separated frequency bands. High frequency gravitational waves (HFGWs), typically above 10~kHz, remain largely unexplored~\cite{Aggarwal:2020olq}. Potential sources include exotic compact objects~\cite{Liebling:2012fv,Cardoso:2019rvt,Palenzuela:2017kcg,Giudice:2016zpa,Banks:2023eym}, primordial black holes~\cite{Wang:2019kaf,Anantua:2008am,Dolgov:2011cq,Dong:2015yjs}, phase transitions~\cite{Hasegawa:2019amx,Croon:2024mde,Kamionkowski:1993fg,Hindmarsh:2013xza,Ai:2025fqw}, inflationary and post-inflationary dynamics~\cite{Easther:2006vd,Dufaux:2007pt,Lozanov:2019ylm}, and cosmic strings~\cite{Servant:2023tua}. A primordial thermal graviton background would also test the thermal history of the early Universe and its compatibility with inflation~\cite{Vagnozzi:2022qmc}. The weakness of the gravitational coupling makes detection challenging. Interferometric searches~\cite{Holometer:2016qoh} and developments in low noise acoustic resonators~\cite{Goryachev:2014nna} illustrate efforts to access this regime.

The inverse Gertsenshtein effect provides an electromagnetic approach: a gravitational wave can convert into a photon of the same frequency in an external magnetic field~\cite{Gertsenshtein1961,Boccaletti1970,Zeldovich1974}. This coupling motivates laboratory magnetic conversion searches~\cite{Ejlli:2019bqj} and detector concepts using axion haloscopes~\cite{Domcke:2022rgu} and microwave cavities~\cite{Berlin:2021txa}. Conversion in planetary magnetic fields~\cite{Liu:2023mll,Tian:2026jqy}, neutron star fields~\cite{Ito:2023fcr,Hong:2024ofh}, and Galactic and extragalactic fields~\cite{Ito:2023nkq,Pshirkov:2009sf,Domcke:2020yzq,Gupta:2026scx,Lecce:2026cqo} extends the same idea to astrophysical environments. The propagation can be described as graviton--photon mixing in the external field~\cite{Raffelt:1987im}. In an ionized medium, plasma dispersion induces a phase mismatch $|\Delta|=\omega_{\rm pl}^2/(2\omega)$, where $\omega_{\rm pl}$ is the plasma frequency and $\omega$ is the wave angular frequency. In a uniform magnetic field, coherent accumulation is lost for propagation lengths much greater than $\Delta^{-1}$, suppressing conversion over astrophysical distances~\cite{Pshirkov:2009sf,Domcke:2020yzq}.

Galactic magnetic field models contain both regular and turbulent components~\cite{Jansson:2012rt,Planck:2016gdp,Unger:2023lob}, and the contribution of turbulence to graviton--photon conversion has previously been considered~\cite{Lella:2024hfgw}. The turbulent field supplies spatial Fourier modes that can compensate the plasma-induced mismatch. A mode with longitudinal wave number $k_z\simeq\Delta$ contributes coherently to the conversion amplitude, so the conversion probability depends on the magnetic power projected along the propagation direction. The central question is how this projected power determines the ensemble-averaged probability and its dependence on frequency and propagation length. Related effects of phase coherence and magnetic structure arise in photon--pseudoscalar conversion~\cite{VanBibber:1987rq,Seong:2023ran,Fiorillo:2026rgo}.

In this work, we derive the projected magnetic spectrum and the ensemble-averaged graviton--photon conversion probability for an isotropic turbulent field with a power law spectrum~\cite{Kolmogorov1941a,Kolmogorov1941b,Addazi:2024kbq,Chiba:2025odu}. In the inertial range regime $\kL\ll\Delta\ll\kH$, where $\kL$ and $\kH$ denote the lower and upper wave number cutoffs of the turbulent spectrum, the probability grows linearly with the magnetized path length $L$ and scales as $\langle P_{g\gamma}\rangle\propto Lf^\beta$, with $\beta=5/3$ for Kolmogorov-like turbulence~\cite{Kolmogorov1941a}. A low mismatch plateau and a high mismatch $\Delta^{-2}$ tail delimit this regime. As an illustrative application, photon background measurements~\cite{Biteau:2025mm} and the Khaire--Srianand extragalactic background light synthesis model~\cite{Khaire:2018mvl} yield characteristic strain limits up to about two orders of magnitude stronger than the previous Galactic conversion bounds of Ref.~\cite{Lella:2024hfgw} in the inertial range phase-matching window. Our results identify Galactic magnetic turbulence as a natural phase-matching medium, providing a mechanism to enhance graviton--photon conversion and improve the sensitivity of electromagnetic searches for high frequency gravitational waves.

{\it Graviton--photon conversion in turbulent Galactic fields.}~~Graviton--photon conversion in an external magnetic field follows from the interaction~\cite{Gertsenshtein1961,Boccaletti1970}
\begin{equation}
    {\cal L}_{\rm int}
    =\frac{\kappa}{2}h_{\mu\nu}T_{\rm em}^{\mu\nu},
\label{eq:graviton_em_interaction}
\end{equation}
where $h_{\mu\nu}$ describes metric perturbations, $T_{\rm em}^{\mu\nu}$ is the electromagnetic energy--momentum tensor, and $\kappa=\sqrt{16\pi G}$. We use natural units, $\hbar=c=1$, and take the propagation direction to be the $z$ axis, with $s$ denoting distance along the path. In the weak conversion limit, neglecting graviton depletion and photon absorption, the polarization averaged probability is determined by the transverse conversion amplitudes
\begin{equation}
    {\cal A}_{g\gamma_j}(\Delta)=
    \frac{\kappa}{2}\int_0^L \dd s\,B_j(s)e^{-i\Delta s},
\label{eq:phase_definition_full}
\end{equation}
\color{black}
with $\Delta\equiv|\Delta_{\rm pl}|=\wpl^2/(2\omega)$ denoting the magnitude of the plasma-induced wave number mismatch between the graviton and photon. Here $\Delta_{\rm pl}$ is the plasma contribution to the photon wave number, $\wpl$ is the electron plasma frequency, and $\omega=2\pi f$ is the wave angular frequency. The index $j=x,y$ labels magnetic components transverse to the line of sight. We assume approximately constant electron density and plasma-dominated photon dispersion. For an unpolarized incident GW, averaging over the two incoming graviton polarizations and summing over the outgoing photon polarizations gives $P_{g\gamma}=\sum_{j=x,y}|{\cal A}_{g\gamma_j}|^2$. The polarization normalization is derived in Supplemental Eqs.~\eqref{eq:sup_mixing}--\eqref{eq:sup_polarization_norm}. Eq.~\eqref{eq:phase_definition_full} shows that conversion samples the transverse magnetic field over the finite path at the longitudinal wave number selected by $\Delta$.

\color{black}
For a uniform transverse field, $B_\perp^2=B_x^2+B_y^2$, direct integration gives $P^{\rm uni}_{g\gamma}(\Delta)=\kappa^2B_\perp^2\Delta^{-2}\sin^2(\Delta L/2)$~\cite{Lella:2024hfgw}. When $\Delta L\ll1$, contributions from the entire path add coherently and $P^{\rm uni}_{g\gamma}\simeq\kappa^2B_\perp^2L^2/4$. When $\Delta L\gg1$, the accumulated plasma phase causes destructive interference between contributions from different locations, leaving an oscillatory probability with an envelope proportional to $\Delta^{-2}$~\cite{Pshirkov:2009sf,Domcke:2020yzq}.

A spatially varying magnetic field can compensate this phase mismatch. For a magnetic Fourier mode with wave vector $\bk$, the phase along the line of sight is $e^{ik_zs}$, where $k_z=\bk\cdot\bn$ and $\bn$ is the propagation direction. Its contribution to Eq.~\eqref{eq:phase_definition_full} is proportional to $\int_0^L\dd s\,e^{i(k_z-\Delta)s}$ and remains coherent when $|k_z-\Delta|L\lesssim1$. Thus the finite path selects longitudinal modes within a wave number interval of order $L^{-1}$ around $k_z=\Delta$. Turbulence supplies such modes without changing the plasma dispersion relation. We denote the probability for a fixed magnetic realization by $P_{g\gamma}$ and its average over realizations by $\langle P_{g\gamma}\rangle$. The latter is determined by the projected magnetic spectrum evaluated with the finite path response, as developed below.

We model the turbulent magnetic field as zero mean, statistically homogeneous, isotropic and divergence free, with rms strength $B_0^2=\langle|\bB|^2\rangle$. Its three dimensional spectrum is $\mathcal P_B(k)\propto k^{-\beta-2}$ for $\kL<k<\kH$, with zero power outside this range. The spectral index satisfies $\beta>1$, with $\beta=5/3$ giving Kolmogorov scaling. A similar slope occurs in the perpendicular cascade of Goldreich--Sridhar turbulence, whose anisotropy is not included in the present isotropic model~\cite{Goldreich:1995}.

The ensemble average of the squared conversion amplitude involves the magnetic two point correlation $\langle B_j(s)B_j(s')\rangle$. Statistical homogeneity makes this correlation depend only on the separation $s-s'$. Fourier transforming the three dimensional magnetic correlation and integrating over transverse wave numbers gives
\begin{equation}
\begin{aligned}
\langle B_j(s)B_j(s')\rangle
={}&B_0^2\int_0^{\kH}\dd k_z\,\Psi(k_z)\\
&\times\cos[k_z(s-s')],\qquad j=x,y.
\end{aligned}
\label{eq:magnetic_correlation}
\end{equation}
\color{black}
Here $\Psi(k_z)$ is the one sided projected magnetic power spectrum of a single transverse component. Its projection integral and explicit finite cutoff expression are given in Supplemental Eqs.~\eqref{eq:psidef} and \eqref{eq:sup_psi}, respectively. The cosine combines the equal contributions at $k_z$ and $-k_z$ for the same component correlation of a real, statistically homogeneous field.
\color{black}
For an isotropic field, each Cartesian component carries one third of the total variance, so $\int_0^{\kH}\dd k_z\,\Psi(k_z)=1/3$. The projected spectrum is obtained by integrating $\mathcal P_B(k)$ over transverse wave numbers at fixed $k_z$, with the tensor weight imposed by $\bk\cdot\bB_{\bk}=0$. Thus it includes all allowed total wave numbers $k=(k_\perp^2+k_z^2)^{1/2}$ sharing that longitudinal component. The magnetic components entering conversion remain those transverse to the propagation direction $\bn$. Although the three dimensional spectrum vanishes below $\kL$, its projection generally extends down to $k_z=0$.

Combining Eq.~\eqref{eq:magnetic_correlation} with Eq.~\eqref{eq:phase_definition_full} gives the finite path probability
\begin{equation}
\begin{aligned}
\langle P_{g\gamma}(\Delta)\rangle
={}&\frac{\kappa^2B_0^2}{4}\int_0^{\kH}\dd k_z\,\Psi(k_z)\\
&\times\bigl[K^2(k_z-\Delta)+K^2(k_z+\Delta)\bigr],
\end{aligned}
\label{eq:p_finite_path}
\end{equation}
\color{black}
The derivation from the magnetic correlation is given in Supplemental Eq.~\eqref{eq:sup_window}. Here $K(x)=2\sin(xL/2)/x$, with $K(0)=L$, is the response of the finite propagation interval. The squared response $K^2(x)$ has peak height proportional to $L^2$ and width proportional to $L^{-1}$. When phase-matched modes are present, $\kL L\gg1$, and the spectrum varies slowly across this width, the leading phase-matched contribution is
\color{black}
\begin{equation}
    \langle P_{g\gamma}(\Delta)\rangle
    \simeq\frac{\pi}{2}\kappa^2B_0^2L\,\Psi(\Delta) .
\label{eq:p_res_full}
\end{equation}
The linear growth with $L$ follows from integrating the coherent response over its narrowing wave number interval. Eq.~\eqref{eq:p_res_full} thus connects the conversion probability directly to the available phase-matched magnetic power.

For a broad inertial range, $\kH\gg\kL$, the projected spectrum and the finite path response yield three asymptotic regimes:

\begin{equation}
\langle P_{g\gamma}\rangle \simeq
\begin{cases}
\displaystyle
\kappa^2B_0^2L\,{\cal C}_0,
& \Delta\ll\kL,\\[1.0ex]
\displaystyle
\kappa^2B_0^2L\,{\cal C}_\beta
(\kL/\Delta)^\beta,
& \kL\ll \Delta\ll\kH,\\[1.2ex]
\displaystyle
\frac{\kappa^2B_0^2}{3\Delta^2},
& \Delta\gg\kH .
\end{cases}
\label{eq:asymptotic_full}
\end{equation}
\color{black}
where ${\cal C}_0=\pi(\beta-1)/(8\beta\kL)$ and ${\cal C}_\beta=\pi(\beta^2-1)/[4\beta(\beta+2)\kL]$. For $\Delta\ll\kL$, the projected spectrum is nearly constant, giving a low mismatch plateau. Phase matching remains possible here because modes with total wave number $k\ge\kL$ can have longitudinal components $k_z=\Delta<\kL$. As $\Delta$ increases through the inertial range, the projected magnetic power at the matching wave number decreases, and the ensemble-averaged conversion probability follows the corresponding spectral power law. For $\Delta\gg\kH$, no turbulent mode can satisfy longitudinal phase matching, and the phase-averaged finite path probability approaches the high mismatch $\Delta^{-2}$ tail in the last line of Eq.~\eqref{eq:asymptotic_full}. Because $\Delta\propto f^{-1}$, the inertial range probability scales as $\langle P_{g\gamma}\rangle\propto f^\beta$. \rev{The inertial range branch describes the enhancement through phase matching and connects its frequency dependence directly to the turbulent spectrum. The low mismatch plateau and the high mismatch tail delimit the frequency interval over which this inertial range scaling applies.} The plateau and inertial range branch follow from Supplemental Eqs.~\eqref{eq:sup_psi} and \eqref{eq:sup_res_probability}. The high mismatch tail instead follows from the phase-averaged finite path response, as explained before Supplemental Eq.~\eqref{eq:sup_asymptotic}.

\color{black}

\begin{figure}[ht]
    \centering
    \includegraphics[width=1\linewidth,height=0.71\linewidth,keepaspectratio]{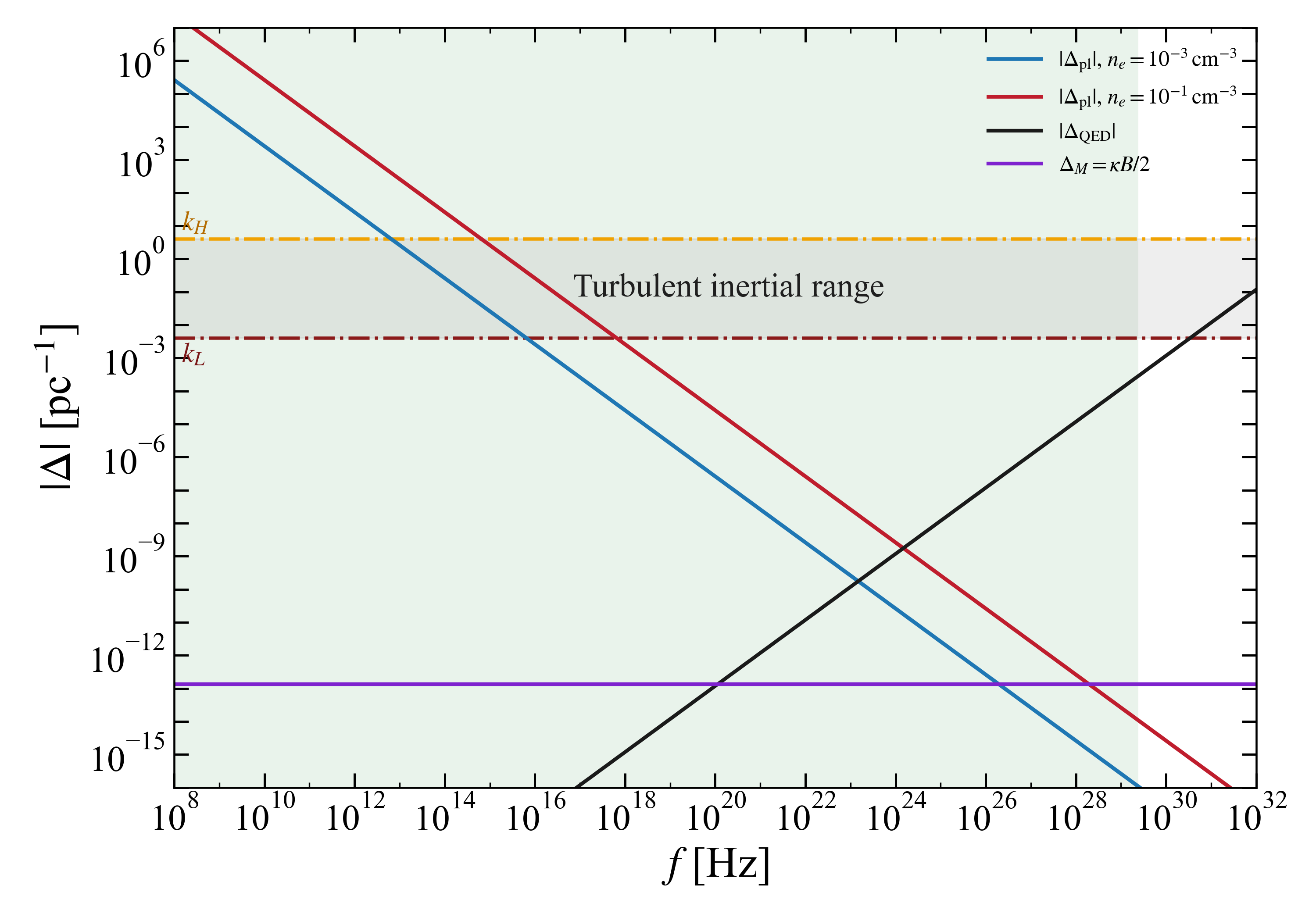}
    \caption{Momentum mismatch and coupling scales used in the benchmark calculation. Colored curves show $|\Delta_{\rm pl}|$ for $\nelec=10^{-3}\,{\rm cm}^{-3}$ and $\nelec=10^{-1}\,{\rm cm}^{-3}$, using the same density colors as Figs.~\ref{fig:prob_full} and \ref{fig:hc_full}. The dark curve is $|\Delta_{\rm QED}|$, and the purple line is the magnetic mixing scale $\Delta_M$. The brown and orange horizontal lines mark $\kL$ and $\kH$, respectively, and the grey band between them is the turbulent inertial range. The pale green background indicates the electromagnetic astronomy band, from radio frequencies to PeV photons. \rev{The figure uses the inverse wave number convention $\kL^{-1}=250\,{\rm pc}$.}}
    \label{fig:scales_full}
\end{figure}

For a chosen outer scale and inertial range width, the boundaries of the inertial range phase-matching window are obtained from $\Delta=\kH$ and $\Delta=\kL$. Table~\ref{tab:benchmarks} lists the magnetic parameters adopted in the benchmark calculation, but is not intended as a complete halo or disk model. In particular, $\kL^{-1}=250\,{\rm pc}$ should be read as an outer scale benchmark. \rev{Faraday rotation studies indicate that the Galactic magneto-ionized medium can have much smaller outer scales in spiral arms and larger scales in inter-arm regions~\cite{Haverkorn:2008outer}.} Changing $\kL$ shifts the inertial range phase-matching window and rescales the low-mismatch plateau, so the numerical bounds below should be read together with this benchmark choice. \rev{We define $L_{\rm out}=\kL^{-1}$, so $\kL=1/(250\,{\rm pc})$. The quantity $2\pi/\kL$ is the associated spatial wavelength, not the tabulated inverse wave number. On the plateau, $\langle P_{g\gamma}\rangle\propto L_{\rm out}$ and $h_c^{\rm lim}\propto L_{\rm out}^{-1/2}$. Changing the outer scale shifts the inertial range phase-matching window and redistributes the power sampled by the mismatch. The enhancement must therefore be evaluated together with the adopted cutoffs.} Figure~\ref{fig:scales_full} shows how the same turbulent wave number band is sampled by two benchmark plasma densities with inertial-range phase-matching windows $6.52\times10^{12}\lesssim f\lesssim6.52\times10^{15}\,\mathrm{Hz}$ for $n_e=10^{-3}\,\mathrm{cm}^{-3}$ and $6.52\times10^{14}\lesssim f\lesssim6.52\times10^{17}\,\mathrm{Hz}$ for $n_e=10^{-1}\,\mathrm{cm}^{-3}$.

\begin{table}[!htb]
\centering
\begin{tabular*}{\columnwidth}{@{\extracolsep{\fill}} lccc}
\hline
     &$|B|$, $B_0$[$\mu$G]&$k_L^{-1}$[pc]&$L$[kpc]\\
\hline
\qquad Coherent \cite{Unger:2023lob}
    & 0.15 & --- & 20 \\
\qquad Turbulent \cite{Jansson:2012rt,Planck:2016gdp}
    & 5 & 250 & 4 \\
\hline
\end{tabular*}
\caption{Adopted magnetic field parameters used for the numerical
illustration. The coherent entry denotes an effective transverse
component, while $B_0$ is the rms turbulent field. For the turbulent benchmarks, we adopt $\kH/\kL=10^3$ and compare two uniform electron densities, $\nelec=10^{-3}\,{\rm cm}^{-3}$ and $\nelec=10^{-1}\,{\rm cm}^{-3}$, keeping the magnetic parameters fixed.}
\label{tab:benchmarks}
\end{table}

\begin{figure}[htbp]
    \centering
    \includegraphics[width=1\linewidth,height=0.71\linewidth,keepaspectratio]{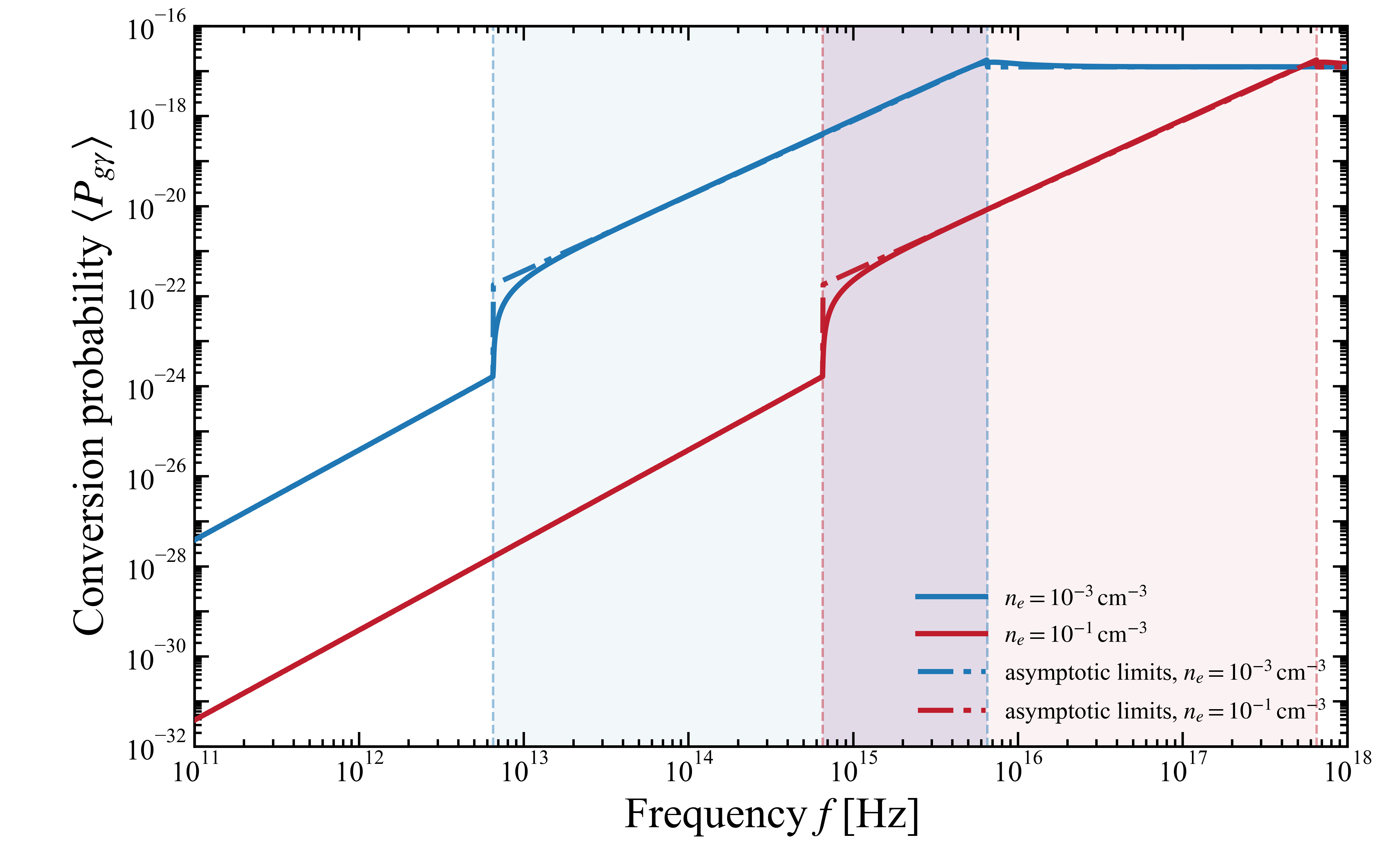}
    \caption{Ensemble-averaged graviton--photon conversion probability over $10^{11}$--$10^{18}\,{\rm Hz}$ for two electron densities. The magnetic parameters are fixed to the benchmark values of Table~\ref{tab:benchmarks}. \rev{Only the electron density $\nelec$ differs between the two curves.} Solid curves show the finite path estimate, while dash dotted curves show the asymptotic limits of Eq.~\eqref{eq:asymptotic_full}. Light shaded bands mark the inertial range phase-matching intervals $\kL<\Delta<\kH$. \rev{Their overlap is the frequency range in which both benchmarks sample the inertial range, and dashed vertical boundaries mark the corresponding crossings.} \rev{Within these intervals, the rise as $f^\beta$ and the path length accumulation follow from matching the plasma phase to longitudinal turbulent modes.} \rev{The figure uses the inverse wave number convention $\kL^{-1}=250\,{\rm pc}$.}}
    \label{fig:prob_full}
\end{figure}

For the probability illustration in Fig.~\ref{fig:prob_full}, we keep the magnetic parameters fixed and vary only $\nelec$. This isolates the role of the plasma mismatch rather than providing a full environment by environment forecast. A denser plasma shifts the same magnetic wave number interval to higher photon frequency because $\Delta\propto \nelec/f$. Inside the inertial range phase-matching window, the dominant contribution comes from modes with longitudinal wave number $k_z\simeq\Delta$, and the middle line of Eq.~\eqref{eq:asymptotic_full} scales as $B_0^2L/\kL$ times a spectral factor. \rev{Stronger turbulent fields and longer magnetized paths raise the conversion probability. Larger outer scales raise the low mismatch plateau, but at fixed $\Delta$ inside the inertial range the scaling is $\langle P_{g\gamma}\rangle\propto \kL^{\beta-1}\Delta^{-\beta}$, so changing the outer scale need not increase the probability there.} The index $\beta$ controls the frequency slope. For Kolmogorov-like turbulence, the inertial range curve rises as $f^{5/3}$ until it reaches the low mismatch plateau. Supplemental Fig.~\ref{fig:mhd_beta_scan} shows how this probability changes for MHD motivated benchmark choices of $\beta$. At low frequencies, where $\Delta>\kH$, the turbulent spectrum contains no mode satisfying $k_z=\Delta$. The probability curves and their inertial range slopes characterize the phase-matching mechanism.

{\it Expected Sensitivity.}~~The conversion probability derived above connects a stochastic high frequency GW background to a potentially observable photon signal. For an isotropic GW background with characteristic strain $h_c(f)$, the ensemble-averaged converted photon intensity is $J_{\gamma}^{\rm conv}(f)=f^2h_c^2(f)\langle P_{g\gamma}(f)\rangle/(16G)$, where $J_\gamma\equiv fI_f$ is the intensity per logarithmic frequency and per steradian, in natural units. Requiring this contribution not to exceed an allowed photon intensity $J_\gamma^{\rm lim}(f)$ gives
\begin{equation}
    h_c(f)\lesssim
    \left[
    \frac{16G\,J_\gamma^{\rm lim}(f)}
    {f^2\langle P_{g\gamma}(f)\rangle}
    \right]^{1/2}.
\label{eq:hc_full}
\end{equation}
Here we use the finite path ensemble-averaged probability of Eq.~\eqref{eq:p_finite_path} for the Galactic benchmarks introduced above. Thus, at fixed photon intensity and frequency, a larger conversion probability gives a tighter strain limit, with $h_c^{\rm lim}\propto\langle P_{g\gamma}\rangle^{-1/2}$. The intensity convention and derivation are given in Supplemental Eqs.~\eqref{eq:sup_photon_density}--\eqref{eq:sup_hc_bound_avg}.

\begin{figure}[ht]
    \centering
    \includegraphics[width=1\linewidth,height=0.6\linewidth]{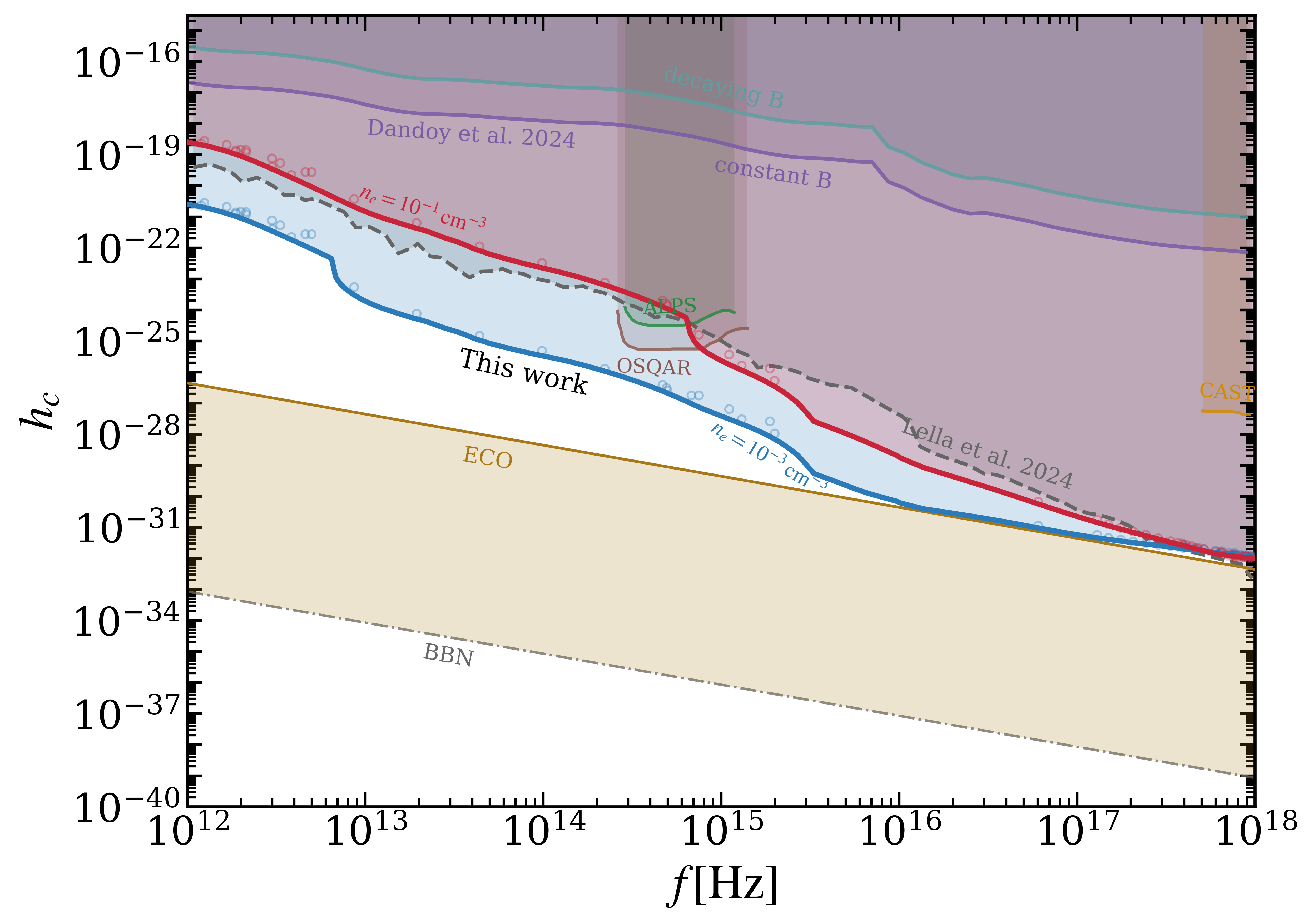}
    \caption{Characteristic strain limits obtained by applying Eq.~\eqref{eq:hc_full} to photon background data and an EBL synthesis model using the ensemble-averaged conversion probability, compared with published high frequency GW constraints over the same $10^{12}$--$10^{18}\,{\rm Hz}$ interval. Open circles show measurement points converted with the conservative pointwise 95\% upper limit described in the Supplemental Material. \rev{Open triangles show reported upper limits whose confidence levels are not homogenized.} Thick blue and red curves show the limits obtained by requiring the converted photon intensity to remain below the Khaire--Srianand EBL model for the two density benchmarks. Thin comparison curves show the overlapping portions of Galactic neutron star bounds~\cite{Dandoy:2024oqg}, laboratory conversion limits from ALPS, OSQAR and CAST~\cite{Ejlli:2019bqj}, the gold ECO line is adapted from the upper envelope of coherent exotic compact object signals in Ref.~\cite{Aggarwal:2020olq}. The shading down to the BBN reference is illustrative and does not denote an exclusion region. }
    \label{fig:hc_full}
\end{figure}

We apply Eq.~\eqref{eq:hc_full} to the photon background measurements and reported upper limits assembled in the public multi-messenger extragalactic spectrum compilation~\cite{Biteau:2025mm}. Its tabulated quantity $\nu I_\nu$ is precisely $J_\gamma$ for $\nu=f$. For a measurement $J_i$ with upper uncertainty $\sigma_{i,+}$, we take $J_\gamma^{\rm lim}=J_i+1.96\sigma_{i,+}$, corresponding to the conservative pointwise 95\% prescription in Supplemental Eqs.~\eqref{eq:sup_profile_chi2}--\eqref{eq:sup_meas_limit}. This prescription allows an unknown nonnegative astrophysical contribution rather than attributing the measured background to conversion. Reported upper limits are used at their original confidence levels. Galaxy count lower limits do not provide upper bounds on an additional photon component and are therefore excluded. The resulting strain limits use the ensemble-mean probability, and the scatter between magnetic field realizations is examined separately in the Supplemental Material.

The extragalactic background light (EBL) synthesis model of Khaire and Srianand~\cite{Khaire:2018mvl} provides a continuous comparison by setting $J_\gamma^{\rm lim}(f)=J_{\rm EBL}(f)$. These model based curves require the converted intensity to remain below the predicted EBL intensity and are not assigned the statistical confidence level of the measurement based limits. Figure~\ref{fig:hc_full} shows both results alongside Galactic neutron star bounds~\cite{Dandoy:2024oqg}, laboratory graviton--photon conversion limits~\cite{Ejlli:2019bqj}, and the previous Galactic conversion bound of Ref.~\cite{Lella:2024hfgw}. The BBN curve provides a cosmological energy density reference. The two density benchmarks illustrate how $\Delta\propto n_e/f$ shifts the frequency range that samples the turbulent inertial range, while $\Psi(\Delta)$ sets the magnetic power available for phase matching. For the adopted Galactic parameters, the strain limits in this range improve by up to about two orders of magnitude over the previous Galactic conversion bound. This comparison illustrates how turbulence enhanced conversion can increase the sensitivity of photon background searches to high frequency gravitational waves.

\color{black}
{\it Conclusion.}~~\rev{We have identified Galactic magnetic turbulence as a natural phase-matching medium for graviton--photon conversion. Its spatial fluctuations supply the longitudinal wave numbers needed to compensate plasma dispersion, allowing matched modes to contribute coherently over the propagation path. The resulting ensemble-averaged probability is determined by the projected magnetic spectrum: in the long path limit, it grows linearly with path length in the phase-matched regime and follows $f^\beta$ when the selected wave number lies within the turbulent inertial range. This relation connects the structure of the Galactic magnetic field to the frequency dependence of the converted photon signal. As an application, photon background observations yield characteristic strain limits up to about two orders of magnitude stronger than previous Galactic conversion bounds in the relevant frequency range for the adopted benchmarks. Our results establish a mechanism through which Galactic turbulence enhances the photon signal from high frequency gravitational waves, providing a theoretical basis for searching for these waves in electromagnetic observations.}
\color{black}

{\it Acknowledgments.}~~L.W. and Y.Z. are supported by the National Natural Science Foundation of China (NNSFC) under grant No. 12335005. B.Z. is supported by the NSFC under Grant No. 12275232 and was the Shandong Province YoungScientists Fund Project (classA) (No. ZR2025QA20).

\renewcommand{\refname}{References}
\bibliography{references}

@article{Raffelt:1987im,
  author = {Raffelt, G. G. and Stodolsky, L.},
  title = {Mixing of the photon with low-mass particles},
  journal = {Phys. Rev. D},
  volume = {37},
  pages = {1237},
  year = {1988},
  doi = {10.1103/PhysRevD.37.1237}
}

@article{Fiorillo:2026rgo,
    author = "Fiorillo, Damiano F. G. and Gil Muyor, {\'A}ngel and Raffelt, Georg G. and Vitagliano, Edoardo",
    title = "{Magnetic Turbulence Boosts Supernova Signals of Axion-Photon Conversion}",
    eprint = "2605.15261",
    archivePrefix = "arXiv",
    primaryClass = "hep-ph",
    doi = "10.1103/wdbg-66y1",
    journal = "Phys. Rev. Lett.",
    volume = "137",
    number = "12",
    pages = "121002",
    year = "2026"
}

@article{Jansson:2012rt,
  author = {Jansson, Ronnie and Farrar, Glennys R.},
  title = {A new model of the Galactic magnetic field},
  journal = {Astrophys. J.},
  volume = {757},
  pages = {14},
  year = {2012},
  doi = {10.1088/0004-637X/757/1/14},
  eprint = {1204.3662},
  archivePrefix = {arXiv},
  primaryClass = {astro-ph.GA}
}

@article{Planck:2016gdp,
  author = {{Planck Collaboration}},
  title = {Planck intermediate results. XLII. Large-scale Galactic magnetic fields},
  journal = {Astron. Astrophys.},
  volume = {596},
  pages = {A103},
  year = {2016},
  doi = {10.1051/0004-6361/201528033},
  eprint = {1601.00546},
  archivePrefix = {arXiv},
  primaryClass = {astro-ph.GA}
}

@article{Haverkorn:2008outer,
  author = {Haverkorn, M. and Brown, J. C. and Gaensler, B. M. and McClure-Griffiths, N. M.},
  title = {The outer scale of turbulence in the magneto-ionized Galactic interstellar medium},
  journal = {Astrophys. J.},
  volume = {680},
  pages = {362},
  year = {2008},
  doi = {10.1086/587165},
  eprint = {0802.2740},
  archivePrefix = {arXiv}
}

@misc{Biteau:2025mm,
  author = {Biteau, Jonathan},
  title = {The Multi-Messenger Extragalactic Spectrum},
  howpublished = {GitLab repository},
  url = {https://gitlab.com/jbiteau.pro/the_multimessenger_extragalactic_spectrum},
  note = {Zenodo doi:10.5281/zenodo.1468852; accessed July 8, 2026},
  year = {2025}
}

@article{Iroshnikov:1964,
  author = {Iroshnikov, R. S.},
  title = {Turbulence of a conducting fluid in a strong magnetic field},
  journal = {Sov. Astron.},
  volume = {7},
  pages = {566},
  year = {1964}
}

@article{Kraichnan:1965,
  author = {Kraichnan, Robert H.},
  title = {Inertial-range spectrum of hydromagnetic turbulence},
  journal = {Phys. Fluids},
  volume = {8},
  pages = {1385},
  year = {1965},
  doi = {10.1063/1.1761412}
}

@article{Goldreich:1995,
  author = {Goldreich, Peter and Sridhar, S.},
  title = {Toward a theory of interstellar turbulence. II. Strong Alfvenic turbulence},
  journal = {Astrophys. J.},
  volume = {438},
  pages = {763},
  year = {1995},
  doi = {10.1086/175121}
}

@article{Lithwick:2001,
  author = {Lithwick, Y. and Goldreich, P.},
  title = {Compressible magnetohydrodynamic turbulence in interstellar plasmas},
  journal = {Astrophys. J.},
  volume = {562},
  pages = {279},
  year = {2001},
  doi = {10.1086/323470},
  eprint = {astro-ph/0106425},
  archivePrefix = {arXiv}
}

@article{Cho:2002,
  author = {Cho, Jungyeon and Lazarian, A.},
  title = {Compressible sub-Alfvenic MHD turbulence in low-beta plasmas},
  journal = {Phys. Rev. Lett.},
  volume = {88},
  pages = {245001},
  year = {2002},
  doi = {10.1103/PhysRevLett.88.245001},
  eprint = {astro-ph/0205282},
  archivePrefix = {arXiv}
}

@article{Cho:2003,
  author = {Cho, Jungyeon and Lazarian, A.},
  title = {Compressible magnetohydrodynamic turbulence: mode coupling, scaling relations, anisotropy, viscosity-damped regime and astrophysical implications},
  journal = {Mon. Not. Roy. Astron. Soc.},
  volume = {345},
  pages = {325},
  year = {2003},
  doi = {10.1046/j.1365-8711.2003.06941.x},
  eprint = {astro-ph/0301062},
  archivePrefix = {arXiv}
}

@article{Ejlli:2019bqj,
  author = {Ejlli, Aldo and Ejlli, Damian and Cruise, Adrian Mike and Pisano, Giampaolo and Grote, Hartmut},
  title = {Upper limits on the amplitude of ultra-high-frequency gravitational waves from graviton to photon conversion},
  journal = {Eur. Phys. J. C},
  volume = {79},
  number = {12},
  pages = {1032},
  year = {2019},
  doi = {10.1140/epjc/s10052-019-7542-5},
  eprint = {1908.00232},
  archivePrefix = {arXiv},
  primaryClass = {gr-qc}
}

@misc{Dandoy:2024oqg,
  author = {Dandoy, Virgile and Bert{\'o}lez-Mart{\'i}nez, Toni and Costa, Francesco},
  title = {High Frequency Gravitational Wave Bounds from Galactic Neutron Stars},
  year = {2024},
  eprint = {2402.14092},
  archivePrefix = {arXiv},
  primaryClass = {gr-qc}
}

@misc{Lella:2024hfgw,
  author = {Lella, Alessandro and Calore, Francesca and Carenza, Pierluca and Mirizzi, Alessandro},
  title = {Constraining gravitational-wave backgrounds from conversions into photons in the Galactic magnetic field},
  year = {2024},
  eprint = {2406.17853},
  archivePrefix = {arXiv},
  primaryClass = {hep-ph}
}

@article{Domcke:2022rgu,
    author = "Domcke, Valerie and Garcia-Cely, Camilo and Rodd, Nicholas L.",
    title = "{Novel Search for High-Frequency Gravitational Waves with Low-Mass Axion Haloscopes}",
    eprint = "2202.00695",
    archivePrefix = "arXiv",
    primaryClass = "hep-ph",
    reportNumber = "DESY-22-017, CERN-TH-2022-010",
    doi = "10.1103/PhysRevLett.129.041101",
    journal = "Phys. Rev. Lett.",
    volume = "129",
    number = "4",
    pages = "041101",
    year = "2022"
}

@article{Liu:2023mll,
    author = "Liu, Tao and Ren, Jing and Zhang, Chen",
    title = "{Limits on High-Frequency Gravitational Waves in Planetary Magnetospheres}",
    eprint = "2305.01832",
    archivePrefix = "arXiv",
    primaryClass = "hep-ph",
    doi = "10.1103/PhysRevLett.132.131402",
    journal = "Phys. Rev. Lett.",
    volume = "132",
    number = "13",
    pages = "131402",
    year = "2024"
}

@article{LIGOScientific:2016aoc,
    author = "Abbott, B. P. and others",
    collaboration = "LIGO Scientific, Virgo",
    title = "{Observation of Gravitational Waves from a Binary Black Hole Merger}",
    eprint = "1602.03837",
    archivePrefix = "arXiv",
    primaryClass = "gr-qc",
    reportNumber = "LIGO-P150914",
    doi = "10.1103/PhysRevLett.116.061102",
    journal = "Phys. Rev. Lett.",
    volume = "116",
    number = "6",
    pages = "061102",
    year = "2016"
}

@article{Croon:2024mde,
    author = "Croon, Djuna and Weir, David J.",
    title = "{Gravitational Waves from Phase Transitions}",
    eprint = "2410.21509",
    archivePrefix = "arXiv",
    primaryClass = "hep-ph",
    reportNumber = "HIP-2024-23/TH",
    doi = "10.1080/00107514.2024.2423496",
    journal = "Contemp. Phys.",
    volume = "65",
    pages = "75",
    year = "2024"
}

@article{NANOGrav:2023gor,
    author = "Agazie, Gabriella and others",
    collaboration = "NANOGrav",
    title = "{The NANOGrav 15 yr Data Set: Evidence for a Gravitational-wave Background}",
    eprint = "2306.16213",
    archivePrefix = "arXiv",
    primaryClass = "astro-ph.HE",
    doi = "10.3847/2041-8213/acdac6",
    journal = "Astrophys. J. Lett.",
    volume = "951",
    number = "1",
    pages = "L8",
    year = "2023"
}

@article{Xu:2023wog,
    author = "Xu, Heng and others",
    title = "{Searching for the Nano-Hertz Stochastic Gravitational Wave Background with the Chinese Pulsar Timing Array Data Release I}",
    eprint = "2306.16216",
    archivePrefix = "arXiv",
    primaryClass = "astro-ph.HE",
    doi = "10.1088/1674-4527/acdfa5",
    journal = "Res. Astron. Astrophys.",
    volume = "23",
    number = "7",
    pages = "075024",
    year = "2023"
}

@article{Aggarwal:2020olq,
    author = "Aggarwal, Nancy and others",
    title = "{Challenges and opportunities of gravitational-wave searches at MHz to GHz frequencies}",
    eprint = "2011.12414",
    archivePrefix = "arXiv",
    primaryClass = "gr-qc",
    reportNumber = "CERN-TH-2020-185, HIP-2020-28/TH, DESY 20-195, CERN-TH-2020-185, HIP-2020-28/TH, DESY 20-195",
    doi = "10.1007/s41114-021-00032-5",
    journal = "Living Rev. Rel.",
    volume = "24",
    number = "1",
    pages = "4",
    year = "2021"
}

@article{Wang:2019kaf,
    author = "Wang, Sai and Terada, Takahiro and Kohri, Kazunori",
    title = "{Prospective constraints on the primordial black hole abundance from the stochastic gravitational-wave backgrounds produced by coalescing events and curvature perturbations}",
    eprint = "1903.05924",
    archivePrefix = "arXiv",
    primaryClass = "astro-ph.CO",
    reportNumber = "KEK-Cosmo-234, KEK-TH-2113",
    doi = "10.1103/PhysRevD.99.103531",
    journal = "Phys. Rev. D",
    volume = "99",
    number = "10",
    pages = "103531",
    year = "2019",
    note = "[Erratum: Phys.Rev.D 101, 069901 (2020)]"
}

@article{Easther:2006vd,
    author = "Easther, Richard and Giblin, Jr., John T. and Lim, Eugene A.",
    title = "{Gravitational Wave Production At The End Of Inflation}",
    eprint = "astro-ph/0612294",
    archivePrefix = "arXiv",
    doi = "10.1103/PhysRevLett.99.221301",
    journal = "Phys. Rev. Lett.",
    volume = "99",
    pages = "221301",
    year = "2007"
}

@article{Dufaux:2007pt,
    author = "Dufaux, Jean Francois and Bergman, Amanda and Felder, Gary N. and Kofman, Lev and Uzan, Jean-Philippe",
    title = "{Theory and Numerics of Gravitational Waves from Preheating after Inflation}",
    eprint = "0707.0875",
    archivePrefix = "arXiv",
    primaryClass = "astro-ph",
    doi = "10.1103/PhysRevD.76.123517",
    journal = "Phys. Rev. D",
    volume = "76",
    pages = "123517",
    year = "2007"
}

@article{Lozanov:2019ylm,
    author = "Lozanov, Kaloian D. and Amin, Mustafa A.",
    title = "{Gravitational perturbations from oscillons and transients after inflation}",
    eprint = "1902.06736",
    archivePrefix = "arXiv",
    primaryClass = "astro-ph.CO",
    doi = "10.1103/PhysRevD.99.123504",
    journal = "Phys. Rev. D",
    volume = "99",
    number = "12",
    pages = "123504",
    year = "2019"
}

@article{Hasegawa:2019amx,
    author = "Hasegawa, Taiki and Okada, Nobuchika and Seto, Osamu",
    title = "{Gravitational waves from the minimal gauged $U(1)_{B-L}$ model}",
    eprint = "1904.03020",
    archivePrefix = "arXiv",
    primaryClass = "hep-ph",
    reportNumber = "EPHOU-19-004",
    doi = "10.1103/PhysRevD.99.095039",
    journal = "Phys. Rev. D",
    volume = "99",
    number = "9",
    pages = "095039",
    year = "2019"
}

@article{Dolgov:2011cq,
    author = "Dolgov, Alexander D. and Ejlli, Damian",
    title = "{Relic gravitational waves from light primordial black holes}",
    eprint = "1105.2303",
    archivePrefix = "arXiv",
    primaryClass = "astro-ph.CO",
    doi = "10.1103/PhysRevD.84.024028",
    journal = "Phys. Rev. D",
    volume = "84",
    pages = "024028",
    year = "2011"
}

@article{Anantua:2008am,
    author = "Anantua, Richard and Easther, Richard and Giblin, John T.",
    title = "{GUT-Scale Primordial Black Holes: Consequences and Constraints}",
    eprint = "0812.0825",
    archivePrefix = "arXiv",
    primaryClass = "astro-ph",
    doi = "10.1103/PhysRevLett.103.111303",
    journal = "Phys. Rev. Lett.",
    volume = "103",
    pages = "111303",
    year = "2009"
}

@article{Dong:2015yjs,
    author = "Dong, Ruifeng and Kinney, William H and Stojkovic, Dejan",
    title = "{Gravitational wave production by Hawking radiation from rotating primordial black holes}",
    eprint = "1511.05642",
    archivePrefix = "arXiv",
    primaryClass = "astro-ph.CO",
    doi = "10.1088/1475-7516/2016/10/034",
    journal = "JCAP",
    volume = "10",
    pages = "034",
    year = "2016"
}

@article{Domcke:2020yzq,
    author = "Domcke, Valerie and Garcia-Cely, Camilo",
    title = "{Potential of radio telescopes as high-frequency gravitational wave detectors}",
    eprint = "2006.01161",
    archivePrefix = "arXiv",
    primaryClass = "astro-ph.CO",
    reportNumber = "DESY-20-097, CERN-TH-2020-082",
    doi = "10.1103/PhysRevLett.126.021104",
    journal = "Phys. Rev. Lett.",
    volume = "126",
    number = "2",
    pages = "021104",
    year = "2021"
}

@article{Ito:2023nkq,
    author = "Ito, Asuka and Kohri, Kazunori and Nakayama, Kazunori",
    title = "{Gravitational Wave Search through Electromagnetic Telescopes}",
    eprint = "2309.14765",
    archivePrefix = "arXiv",
    primaryClass = "gr-qc",
    reportNumber = "KEK-QUP-2023-0018, KEK-TH-2558, KEK-Cosmo-0327, TU-1205",
    doi = "10.1093/ptep/ptae004",
    journal = "PTEP",
    volume = "2024",
    number = "2",
    pages = "023E03",
    year = "2024"
}

@article{Ito:2023fcr,
    author = "Ito, Asuka and Kohri, Kazunori and Nakayama, Kazunori",
    title = "{Probing high frequency gravitational waves with pulsars}",
    eprint = "2305.13984",
    archivePrefix = "arXiv",
    primaryClass = "gr-qc",
    reportNumber = "KEK-QUP-2023-0011, KEK-TH-2529, KEK-Cosmo-0314, TU-1192",
    doi = "10.1103/PhysRevD.109.063026",
    journal = "Phys. Rev. D",
    volume = "109",
    number = "6",
    pages = "063026",
    year = "2024"
}

@article{Holometer:2016qoh,
    author = "Chou, Aaron S. and others",
    collaboration = "Holometer",
    title = "{MHz Gravitational Wave Constraints with Decameter Michelson Interferometers}",
    eprint = "1611.05560",
    archivePrefix = "arXiv",
    primaryClass = "astro-ph.IM",
    reportNumber = "FERMILAB-PUB-16-449-AE",
    doi = "10.1103/PhysRevD.95.063002",
    journal = "Phys. Rev. D",
    volume = "95",
    number = "6",
    pages = "063002",
    year = "2017"
}

@article{Goryachev:2014nna,
    author = "Goryachev, Maxim and Ivanov, Eugene N. and van Kann, Frank and Galliou, Serge and Tobar, Michael E.",
    title = "{Observation of the Fundamental Nyquist Noise Limit in an Ultra-High $Q$-Factor Cryogenic Bulk Acoustic Wave Cavity}",
    eprint = "1410.4293",
    archivePrefix = "arXiv",
    primaryClass = "physics.ins-det",
    doi = "10.1063/1.4898813",
    journal = "Appl. Phys. Lett.",
    volume = "105",
    pages = "153505",
    year = "2014"
}

@article{Gertsenshtein:1962kfm,
    author = "Gertsenshtein, M. E. and Pustovoit, V. I.",
    title = "{On the detection of low frequency gravitational waves}",
    journal = "Sov. Phys. JETP",
    volume = "16",
    pages = "433--435",
    year = "1962"
}

@article{Hong:2024ofh,
    author = "Hong, Wei and Tao, Zhen-Zhao and He, Peng and Zhang, Tong-Jie",
    title = "{Theoretical Radio Signals from Radio-band Gravitational Waves Converted from the Neutron Star Magnetic Field}",
    eprint = "2412.05338",
    archivePrefix = "arXiv",
    primaryClass = "astro-ph.HE",
    doi = "10.3847/1538-4357/adf19a",
    journal = "Astrophys. J.",
    volume = "990",
    number = "2",
    pages = "156",
    year = "2025"
}

@article{Zeldovich1974,
  author = {Zel'dovich, Y. B.},
  title = {Electromagnetic and gravitational waves in a stationary magnetic field},
  journal = {Soviet Journal of Experimental and Theoretical Physics},
  volume = {38},
  pages = {652},
  year = {1974}
}

@article{Gertsenshtein1961,
  author = {Gertsenshtein, M. E.},
  title = {Wave Resonance of Light and Gravitational Waves},
  journal = {Journal of Experimental and Theoretical Physics},
  volume = {41},
  pages = {113--114},
  year = {1961},
  note = {English translation in \textit{Soviet Physics JETP}, Vol. 14, pp. 84--85 (1962)}
}

@article{Berlin:2021txa,
    author = {Berlin, Asher and Blas, Diego and Tito D'Agnolo, Raffaele and Ellis, Sebastian A. R. and Harnik, Roni and Kahn, Yonatan and Sch{\"u}tte-Engel, Jan},
    title = "{Detecting high-frequency gravitational waves with microwave cavities}",
    eprint = "2112.11465",
    archivePrefix = "arXiv",
    primaryClass = "hep-ph",
    reportNumber = "FERMILAB-PUB-21-724-SQMS-T",
    doi = "10.1103/PhysRevD.105.116011",
    journal = "Phys. Rev. D",
    volume = "105",
    number = "11",
    pages = "116011",
    year = "2022"
}

@article{Boccaletti1970,
  author = {Boccaletti, D. and De Sabbata, V. and Fortini, P. and Gualdi, C.},
  title = {Conversion of photons into gravitons and vice versa in a static electromagnetic field},
  journal = {Il Nuovo Cimento B (1965-1970)},
  volume = {70},
  pages = {129--146},
  year = {1970}
}

@article{EPTA:2023fyk,
    author = "Antoniadis, J. and others",
    collaboration = "EPTA, InPTA:",
    title = "{The second data release from the European Pulsar Timing Array - III. Search for gravitational wave signals}",
    eprint = "2306.16214",
    archivePrefix = "arXiv",
    primaryClass = "astro-ph.HE",
    doi = "10.1051/0004-6361/202346844",
    journal = "Astron. Astrophys.",
    volume = "678",
    pages = "A50",
    year = "2023"
}

@article{Reardon:2023gzh,
    author = "Reardon, Daniel J. and others",
    title = "{Search for an Isotropic Gravitational-wave Background with the Parkes Pulsar Timing Array}",
    eprint = "2306.16215",
    archivePrefix = "arXiv",
    primaryClass = "astro-ph.HE",
    doi = "10.3847/2041-8213/acdd02",
    journal = "Astrophys. J. Lett.",
    volume = "951",
    number = "1",
    pages = "L6",
    year = "2023"
}

@article{Kamionkowski:1993fg,
    author = "Kamionkowski, Marc and Kosowsky, Arthur and Turner, Michael S.",
    title = "{Gravitational radiation from first order phase transitions}",
    eprint = "astro-ph/9310044",
    archivePrefix = "arXiv",
    reportNumber = "IASSNS-HEP-93-44, FERMILAB-PUB-93-235-A",
    doi = "10.1103/PhysRevD.49.2837",
    journal = "Phys. Rev. D",
    volume = "49",
    pages = "2837--2851",
    year = "1994"
}

@article{Hindmarsh:2013xza,
    author = "Hindmarsh, Mark and Huber, Stephan J. and Rummukainen, Kari and Weir, David J.",
    title = "{Gravitational waves from the sound of a first order phase transition}",
    eprint = "1304.2433",
    archivePrefix = "arXiv",
    primaryClass = "hep-ph",
    reportNumber = "HIP-2013-07-TH",
    doi = "10.1103/PhysRevLett.112.041301",
    journal = "Phys. Rev. Lett.",
    volume = "112",
    pages = "041301",
    year = "2014"
}

@article{Pshirkov:2009sf,
    author = "Pshirkov, M. S. and Baskaran, D.",
    title = "{Limits on High-Frequency Gravitational Wave Background from its interplay with Large Scale Magnetic Fields}",
    eprint = "0903.4160",
    archivePrefix = "arXiv",
    primaryClass = "gr-qc",
    doi = "10.1103/PhysRevD.80.042002",
    journal = "Phys. Rev. D",
    volume = "80",
    pages = "042002",
    year = "2009"
}

@article{Palenzuela:2017kcg,
    author = "Palenzuela, Carlos and Pani, Paolo and Bezares, Miguel and Cardoso, Vitor and Lehner, Luis and Liebling, Steven",
    title = "{Gravitational Wave Signatures of Highly Compact Boson Star Binaries}",
    eprint = "1710.09432",
    archivePrefix = "arXiv",
    primaryClass = "gr-qc",
    doi = "10.1103/PhysRevD.96.104058",
    journal = "Phys. Rev. D",
    volume = "96",
    number = "10",
    pages = "104058",
    year = "2017"
}

@article{Giudice:2016zpa,
    author = "Giudice, Gian F. and McCullough, Matthew and Urbano, Alfredo",
    title = "{Hunting for Dark Particles with Gravitational Waves}",
    eprint = "1605.01209",
    archivePrefix = "arXiv",
    primaryClass = "hep-ph",
    reportNumber = "CERN-TH-2016-106",
    doi = "10.1088/1475-7516/2016/10/001",
    journal = "JCAP",
    volume = "10",
    pages = "001",
    year = "2016"
}

@article{Banks:2023eym,
    author = "Banks, Hannah and Grabowska, Dorota M. and McCullough, Matthew",
    title = "{Gravitational wave backgrounds from colliding exotic compact objects}",
    eprint = "2302.07887",
    archivePrefix = "arXiv",
    primaryClass = "gr-qc",
    reportNumber = "CERN-TH-2023-016, IQuS@UW-21-044",
    doi = "10.1103/PhysRevD.108.035017",
    journal = "Phys. Rev. D",
    volume = "108",
    number = "3",
    pages = "035017",
    year = "2023"
}

@article{Cardoso:2019rvt,
    author = "Cardoso, Vitor and Pani, Paolo",
    title = "{Testing the nature of dark compact objects: a status report}",
    eprint = "1904.05363",
    archivePrefix = "arXiv",
    primaryClass = "gr-qc",
    doi = "10.1007/s41114-019-0020-4",
    journal = "Living Rev. Rel.",
    volume = "22",
    number = "1",
    pages = "4",
    year = "2019"
}

@article{Liebling:2012fv,
    author = "Liebling, Steven L. and Palenzuela, Carlos",
    title = "{Dynamical boson stars}",
    eprint = "1202.5809",
    archivePrefix = "arXiv",
    primaryClass = "gr-qc",
    doi = "10.1007/s41114-023-00043-4",
    journal = "Living Rev. Rel.",
    volume = "26",
    number = "1",
    pages = "1",
    year = "2023"
}

@article{Servant:2023tua,
    author = "Servant, G{\'e}raldine and Simakachorn, Peera",
    title = "{Ultrahigh frequency primordial gravitational waves beyond the kHz: The case of cosmic strings}",
    eprint = "2312.09281",
    archivePrefix = "arXiv",
    primaryClass = "hep-ph",
    reportNumber = "DESY-23-202, CERN-TH-2023-226",
    doi = "10.1103/PhysRevD.109.103538",
    journal = "Phys. Rev. D",
    volume = "109",
    number = "10",
    pages = "103538",
    year = "2024"
}

@article{Ai:2025fqw,
    author = "Ai, Wen-Yuan",
    title = "{High-frequency gravitational waves from a first-order phase transitions}",
    eprint = "2508.02794",
    archivePrefix = "arXiv",
    primaryClass = "hep-ph",
    doi = "10.1103/4v2z-fqsx",
    journal = "Phys. Rev. D",
    volume = "113",
    number = "5",
    pages = "056007",
    year = "2026"
}

@article{Miles:2024seg,
    author = "Miles, Matthew T. and others",
    title = "{The MeerKAT Pulsar Timing Array: the first search for gravitational waves with the MeerKAT radio telescope}",
    eprint = "2412.01153",
    archivePrefix = "arXiv",
    primaryClass = "astro-ph.HE",
    doi = "10.1093/mnras/stae2571",
    journal = "Mon. Not. Roy. Astron. Soc.",
    volume = "536",
    number = "2",
    pages = "1489--1500",
    year = "2024"
}

@misc{Gupta:2026scx,
    author = "Gupta, Aman and Majumdar, Pratik and Roy, Sourov and Sarkar, Pratick",
    title = "{High-Frequency Gravitational Wave Constraints from Graviton-Photon Conversion in the M87 Galaxy}",
    eprint = "2604.01290",
    archivePrefix = "arXiv",
    primaryClass = "hep-ph",
    month = "4",
    year = "2026"
}

@article{Vagnozzi:2022qmc,
    author = "Vagnozzi, Sunny and Loeb, Abraham",
    title = "{The Challenge of Ruling Out Inflation via the Primordial Graviton Background}",
    eprint = "2208.14088",
    archivePrefix = "arXiv",
    primaryClass = "astro-ph.CO",
    doi = "10.3847/2041-8213/ac9b0e",
    journal = "Astrophys. J. Lett.",
    volume = "939",
    number = "2",
    pages = "L22",
    year = "2022"
}

@article{VanBibber:1987rq,
    author = "Van Bibber, K. and Dagdeviren, N. R. and Koonin, S. E. and Kerman, A. and Nelson, H. N.",
    title = "{Proposed experiment to produce and detect light pseudoscalars}",
    reportNumber = "SLAC-PUB-4322",
    doi = "10.1103/PhysRevLett.59.759",
    journal = "Phys. Rev. Lett.",
    volume = "59",
    pages = "759--762",
    year = "1987"
}

@article{Khaire:2018mvl,
    author = "Khaire, Vikram and Srianand, Raghunathan",
    title = "{New synthesis models of consistent extragalactic background light over cosmic time}",
    eprint = "1801.09693",
    archivePrefix = "arXiv",
    primaryClass = "astro-ph.GA",
    doi = "10.1093/mnras/stz174",
    journal = "Mon. Not. Roy. Astron. Soc.",
    volume = "484",
    number = "3",
    pages = "4174--4199",
    year = "2019"
}

@article{Kolmogorov1941a,
  author  = {Kolmogorov, A. N.},
  title   = {The local structure of turbulence in incompressible viscous fluid for very large Reynolds numbers},
  journal = {Dokl. Akad. Nauk SSSR},
  volume  = {30},
  pages   = {299},
  year    = {1941},
  note    = {Reprinted in Proc. Roy. Soc. A \textbf{434} (1991) 9}
}

@article{Kolmogorov1941b,
  author  = {Kolmogorov, A. N.},
  title   = {On the degeneration of isotropic turbulence in an incompressible viscous liquid},
  journal = {Dokl. Akad. Nauk SSSR},
  volume  = {32},
  pages   = {16},
  year    = {1941},
  note    = {Reprinted in Proc. Roy. Soc. A \textbf{434} (1991) 15}
}

@article{Tian:2026jqy,
    author = "Tian, Hongliang and Wu, Lei and Yang, Xiaolong and Yuan, Qiang and Zhu, Bin",
    title = "{Search for High-Frequency Gravitational Waves via Geomagnetic Conversion with Radio Telescopes}",
    journal = {},
    eprint = "2606.13642",
    archivePrefix = "arXiv",
    primaryClass = "gr-qc",
    month = "6",
    year = "2026"
}

@article{Seong:2023ran,
    author = "Seong, Hyeonseok and Sun, Chen and Yun, Seokhoon",
    title = "{Axion magnetic resonance: A novel enhancement in axion-photon conversion}",
    eprint = "2308.10925",
    archivePrefix = "arXiv",
    primaryClass = "hep-ph",
    reportNumber = "DESY-23-118, LA-UR-23-29378, CTPU-PTC-23-39",
    doi = "10.1103/PhysRevD.110.015018",
    journal = "Phys. Rev. D",
    volume = "110",
    number = "1",
    pages = "015018",
    year = "2024"
}

@article{Unger:2023lob,
    author = "Unger, Michael and Farrar, Glennys R.",
    title = "{The Coherent Magnetic Field of the Milky Way}",
    eprint = "2311.12120",
    archivePrefix = "arXiv",
    primaryClass = "astro-ph.GA",
    doi = "10.3847/1538-4357/ad4a54",
    journal = "Astrophys. J.",
    volume = "970",
    number = "1",
    pages = "95",
    year = "2024"
}

@article{Addazi:2024kbq,
    author = "Addazi, Andrea and Capozziello, Salvatore and Gan, Qingyu",
    title = "{Resonant graviton{\textendash}photon conversion with stochastic magnetic field in the expanding universe}",
    eprint = "2401.15965",
    archivePrefix = "arXiv",
    primaryClass = "gr-qc",
    doi = "10.1016/j.dark.2025.101844",
    journal = "Phys. Dark Univ.",
    volume = "48",
    pages = "101844",
    year = "2025"
}

@article{Chiba:2025odu,
    author = "Chiba, Wataru and Jinno, Ryusuke and Nomura, Kimihiro",
    title = "{Graviton-photon conversion in stochastic magnetic fields}",
    eprint = "2505.10926",
    archivePrefix = "arXiv",
    primaryClass = "gr-qc",
    reportNumber = "KOBE-COSMO-25-08, KUNS-3048",
    doi = "10.1103/vjj8-xf7k",
    journal = "Phys. Rev. D",
    volume = "112",
    number = "8",
    pages = "083539",
    year = "2025"
}

@article{Lecce:2026cqo,
    author = "Lecce, Francesca and Cand{\`o}n, Francisco R. and Lella, Alessandro and Mirizzi, Alessandro and Vogel, Julia K. and Zanzarella, Tiziano",
    title = "{Galaxy clusters as cosmic lenses for high frequency gravitational-waves}",
    eprint = "2611.13105",
    journal={},
    archivePrefix = "arXiv",
    primaryClass = "hep-ph",
    month = "11",
    year = "2026"
}

\onecolumngrid
\clearpage

\setcounter{page}{1}
\setcounter{section}{0}
\setcounter{subsection}{0}
\renewcommand{\thesection}{\Roman{section}}
\renewcommand{\thesubsection}{\Alph{subsection}}

\setcounter{equation}{0}
\setcounter{figure}{0}
\setcounter{table}{0}
\renewcommand{\theequation}{S.\arabic{equation}}
\renewcommand{\thefigure}{S\arabic{figure}}
\renewcommand{\thetable}{S\arabic{table}}
\renewcommand{\theHequation}{S.\arabic{equation}}
\renewcommand{\theHfigure}{S\arabic{figure}}
\renewcommand{\theHtable}{S\arabic{table}}





\onecolumngrid
\clearpage

\setcounter{page}{1}
\setcounter{equation}{0}
\setcounter{figure}{0}
\setcounter{table}{0}
\setcounter{section}{0}
\setcounter{subsection}{0}
\renewcommand{\theequation}{S.\arabic{equation}}
\renewcommand{\thefigure}{S\arabic{figure}}
\renewcommand{\thetable}{S\arabic{table}}
\renewcommand{\thesection}{\Roman{section}}
\renewcommand{\thesubsection}{\Alph{subsection}}
\newcommand{\ssection}[1]{
    \addtocounter{section}{1}
    \section{\thesection.~~~#1}
    \addtocounter{section}{-1}
    \refstepcounter{section}
}
\newcommand{\ssubsection}[1]{
    \addtocounter{subsection}{1}
    \subsection{\thesubsection.~~~#1}
    \addtocounter{subsection}{-1}
    \refstepcounter{subsection}
}
\newcommand{\fakeaffil}[2]{$^{#1}$\textit{#2}\\}

\setcounter{figure}{0}
\setcounter{table}{0}

\author{Hongliang Tian}
\email{hongliangtian@njnu.edu.cn}
\affiliation{Department of Physics and Institute of Theoretical Physics, Nanjing Normal University, Nanjing, 210023, China}

\author{Lei Wu}
\email{leiwu@njnu.edu.cn}
\affiliation{Department of Physics and Institute of Theoretical Physics, Nanjing Normal University, Nanjing, 210023, China}
\affiliation{Nanjing Key Laboratory of Particle Physics and Astrophysics, Nanjing, 210023, China}

\author{Yang Zhang}
\email{zhangyang2025@htu.edu.cn}
\affiliation{Center for Theoretical Physics, Henan Normal University, Xinxiang 453007, China}

\author{Bin Zhu}
\email{zhubin@mail.nankai.edu.cn}
\affiliation{School of Physics, Yantai University, Yantai 264005, China}

\thispagestyle{empty}
\begin{center}
    \begin{spacing}{1.2}
        \textbf{\large 
            \hypertarget{sm}{Supplemental Material:}
            \\Enhancing Graviton--Photon Conversion through Galactic Magnetic Turbulence
        }
    \end{spacing}
    \par\smallskip
    Hongliang Tian,$^{1}$
    Lei Wu,$^{1,2}$
    Yang Zhang,$^{3}$
    Bin Zhu$^{4}$

    \par
    {\small
        \fakeaffil{1}{Department of Physics and Institute of Theoretical Physics, Nanjing Normal University, Nanjing, 210023, China}
        \fakeaffil{2}{Nanjing Key Laboratory of Particle Physics and Astrophysics, Nanjing, 210023, China} 
        \fakeaffil{3}{Center for Theoretical Physics, Henan Normal University, Xinxiang 453007, China}
        \fakeaffil{4}{School of Physics, Yantai University, Yantai 264005, China} 
    }

\end{center}

\color{black}
This Supplemental Material presents the theoretical framework and uncertainty analysis for graviton-photon conversion in Galactic magnetic turbulence. We first derive the weak-mixing conversion amplitude and the projected magnetic power spectrum, and examine the dependence on the turbulent spectral index. We then describe the finite path calculation and derive the conversion probabilities in the low mismatch, inertial range, and high mismatch regimes. Next, we establish the relation between the photon background intensity and the characteristic strain, and explain the treatment of measurements, reported upper limits, and the EBL model. Finally, we discuss the magnetic realization uncertainty.\\

{\bf Weak-mixing setup.} We use the interaction and coupling convention of the main text. For propagation along $s$, let $\mathbf h=(h_+,h_\times)^T$ and $\mathbf A=(A_x,A_y)^T$. Neglecting absorption and polarization dependent refractive terms, the polarization equations can be written, up to phase conventions, as
\begin{equation}
\left[i\partial_s+\begin{pmatrix}0&\mathsf M^T(s)\\\mathsf M(s)&\Delta_\gamma(s)\mathbf 1_2\end{pmatrix}\right]
\begin{pmatrix}\mathbf h\\\mathbf A\end{pmatrix}=0,
\qquad
\mathsf M(s)=\frac{\kappa}{2}\begin{pmatrix}B_y(s)&B_x(s)\\B_x(s)&-B_y(s)\end{pmatrix}.
\label{eq:sup_mixing}
\end{equation}
Here both diagonal blocks are $2\times2$. For a fixed transverse field direction, a constant polarization basis reduces this system to two independent two state channels. For a varying direction, we retain the matrix. The local mixing scale in Fig.~\ref{fig:scales_full} is $\Delta_M=\kappa B_\perp/2$. Neglecting graviton depletion, the first order transfer amplitudes are determined by
\color{black}
\begin{equation}
    {\cal A}_{g\gamma_j}=
    \frac{\kappa}{2}\int_0^L \dd s\, B_j(s)
    \exp\!\left[-i\int_0^s \dd l\,\Delta_\gamma(l)\right].
\label{eq:sup_amplitude}
\end{equation}
\color{black}
The transfer matrix is $\mathsf T=\bigl(\begin{smallmatrix}\mathcal A_{g\gamma_y}&\mathcal A_{g\gamma_x}\\\mathcal A_{g\gamma_x}&-\mathcal A_{g\gamma_y}\end{smallmatrix}\bigr)$, up to a common phase. For a unit-normalized unpolarized incident GW,
\begin{equation}
P_{g\gamma}=\tfrac12\operatorname{Tr}(\mathsf T^\dagger\mathsf T)
=|\mathcal A_{g\gamma_x}|^2+|\mathcal A_{g\gamma_y}|^2.
\label{eq:sup_polarization_norm}
\end{equation}
This averages over the incoming polarizations and sums over the outgoing polarizations, as in the main text.

\color{black}
For the Galactic benchmarks considered in the main text, the QED vacuum polarization term is negligible compared with the plasma term, so $\Delta_\gamma\simeq\Delta_{\rm pl}$. The main text uses $\Delta=|\Delta_{\rm pl}|$ and the Fourier convention $e^{-i\Delta s}$. \rev{Changing this sign complex conjugates the amplitude and leaves all probabilities unchanged.} For approximately constant electron density, squaring the amplitude and summing over photon polarizations gives
\begin{align}
    P_{g\gamma}(\Delta)=&
    \frac{\kappa^2}{4}\int_0^L\dd s
    \int_0^L\dd s'\,
    \bB_\perp(s)\!\cdot\!\bB_\perp(s')
    \nonumber\\
    &\times\cos[\Delta(s-s')].
\label{eq:sup_prob}
\end{align}
Here $\bB_\perp=(B_x,B_y)$ is the magnetic field transverse to the propagation direction.

\color{black}
{\bf Projected magnetic power spectrum.} Let us consider the general structure of turbulence, which we take to be isotropic and divergence free. The index $\beta$ denotes the inertial range magnetic spectral slope introduced in the main text. \rev{The value $\beta=5/3$ corresponds to Kolmogorov scaling.} The Fourier space magnetic two point correlation is taken to be
\color{black}
\begin{equation}
    \langle B_{i,\bk}B^*_{j,\bk'}\rangle
    =
    \left(\delta_{ij}-\frac{k_i k_j}{k^2}\right)
    (2\pi)^3\delta(\bk-\bk')
    {\cal P}_B(k),
\end{equation}
where $B_{i,\bk}$ is the Fourier coefficient of the $i$th magnetic field component, $\delta(\bk-\bk')$ enforces statistical homogeneity, and the transverse projector enforces $\nabla\cdot\bB=0$. The scalar power spectrum is
\begin{equation}
    {\cal P}_B(k)=
    \frac{\pi^2(\beta-1)B_0^2}
    {\kL^{-\beta+1}-\kH^{-\beta+1}}
    k^{-\beta-2},
    \qquad \kL<k<\kH ,
\end{equation}
\color{black}
and ${\cal P}_B(k)=0$ outside the inertial range. The normalization is chosen such that $\langle|\bB(\mathbf{x})|^2\rangle=B_0^2$. We now project this correlation along a specific line of sight, which we take to be the $z$ axis. Writing $\mathbf{k}=(\mathbf{k}_\perp,k_z)$, the transverse correlation is
\color{black}
\begin{align}
    \langle B_x(z_1)B_x(z_2)\rangle
    =&\int_{-\infty}^{+\infty}\frac{\dd k_z}{2\pi}
    e^{ik_z(z_1-z_2)}
    \int\frac{\dd^2\mathbf{k}_\perp}{(2\pi)^2}
    \left(1-\frac{k_\perp^2\cos^2\phi}{k_\perp^2+k_z^2}\right)
    {\cal P}_B\!\left(\sqrt{k_\perp^2+k_z^2}\right).
\end{align}
Here $\phi$ is the azimuthal angle of $\mathbf{k}_\perp$ measured from the $x$ axis. For a fixed longitudinal wave number $k_z$, all transverse wave numbers $k_\perp$ with total magnitude $k=(k_\perp^2+k_z^2)^{1/2}$ inside the inertial range contribute to the same longitudinal phase. The tensor factor in parentheses is the solenoidal projection. It removes the component of the magnetic field parallel to $\bk$ and leaves only the part that can appear in the transverse mixing matrix.

After the angular integral, this can be written as
\begin{equation}
    \langle B_x(z_1)B_x(z_2)\rangle
    =
    B_0^2\int_0^{\kH}\dd k_z\,\Psi(k_z)
    \cos[k_z(z_1-z_2)] ,
\label{eq:sup_projected_corr}
\end{equation}
where
\begin{equation}
    \Psi(k_z)
    =
    \frac{\beta-1}{\kL^{-\beta+1}-\kH^{-\beta+1}}
    \int\frac{k_\perp\,\dd k_\perp}{2}
    \left[
    1-\frac{k_\perp^2}{2(k_\perp^2+k_z^2)}
    \right]
    (k_\perp^2+k_z^2)^{-\beta/2-1}.
    \label{eq:psidef}
\end{equation}
\color{black}
The integral is restricted to $\kL^2<k_\perp^2+k_z^2<\kH^2$. The lower boundary is the origin of the quantity $\kmin=\max(|k_z|,\kL)$ below. If $|k_z|<\kL$, the longitudinal mode cannot resolve scales larger than the outer scale and the projection begins at $\kL$. If $|k_z|>\kL$, the longitudinal phase itself fixes the smallest allowed total wave number. Evaluating the integral with its finite cutoffs gives
\begin{equation}
\Psi(k_z)=
\frac{\beta-1}{4(\kL^{1-\beta}-\kH^{1-\beta})}
\left[
\frac{\kmin^{-\beta}-\kH^{-\beta}}{\beta}
+\frac{k_z^2}{\beta+2}
\left(\kmin^{-\beta-2}-\kH^{-\beta-2}\right)
\right],
\qquad
\kmin=\max(|k_z|,\kL).
\label{eq:sup_psi}
\end{equation}
This expression applies for $|k_z|\leq\kH$, with zero power outside, and obeys $\int_0^{\kH}\Psi(k_z)\,\dd k_z=1/3$. For the leading broad range asymptotics below, $(\kL^{1-\beta}-\kH^{1-\beta})^{-1}\simeq\kL^{\beta-1}$. The same function appears for $B_y$. The longitudinal component has a different projection factor and does not enter graviton--photon conversion for propagation along $z$. Equation~\eqref{eq:sup_psi} should therefore be read as a one dimensional magnetic power spectrum. It is the amount of transverse turbulent magnetic power available at the longitudinal wave number selected by the plasma phase mismatch.

\color{black}
The scalar index $\beta$ used above is a phenomenological input to the projected, isotropic model. Real magnetohydrodynamic (MHD) turbulence contains Alfven, slow and fast cascades, and these modes need not share the same anisotropy or spectral slope. Strong Alfvenic turbulence is commonly associated with a Kolmogorov-like perpendicular cascade, $\beta=5/3$, in the Goldreich--Sridhar picture~\cite{Goldreich:1995}. In compressible MHD, slow modes are often treated as passively mixed by the Alfvenic cascade, while fast modes can follow a more isotropic and sometimes Iroshnikov--Kraichnan-like scaling~\cite{Iroshnikov:1964,Kraichnan:1965,Lithwick:2001,Cho:2002,Cho:2003}. Our calculation does not decompose the turbulent field into these anisotropic eigenmodes. Instead, we use $\beta$ as a controlled benchmark parameter that redistributes a fixed rms turbulent field $B_0$ over the same inertial range.

Figure~\ref{fig:mhd_beta_scan} shows the finite path numerical probability from Eq.~\eqref{eq:sup_window} for three representative choices, $\beta=3/2$, $5/3$ and $2$. Colors denote the two electron density benchmarks, and line styles denote the spectral index benchmarks. The change in $\beta$ is most visible when the plasma mismatch lies inside the inertial range, $\kL\lesssim\Delta\lesssim\kH$, where Eq.~\eqref{eq:sup_asymptotic} gives $\langle P_{g\gamma}\rangle\propto\Delta^{-\beta}\propto f^\beta$. Smaller $\beta$ leaves relatively more power at high wave number and therefore gives a larger probability at the low frequency side of the inertial range phase matching window. Larger $\beta$ suppresses that part of the curve more strongly, but gives a slightly larger low mismatch plateau through the factor $(\beta-1)/\beta$ in Eq.~\eqref{eq:sup_asymptotic}.

\begin{figure}[!tbp]
    \centering
    \includegraphics[width=0.82\textwidth]{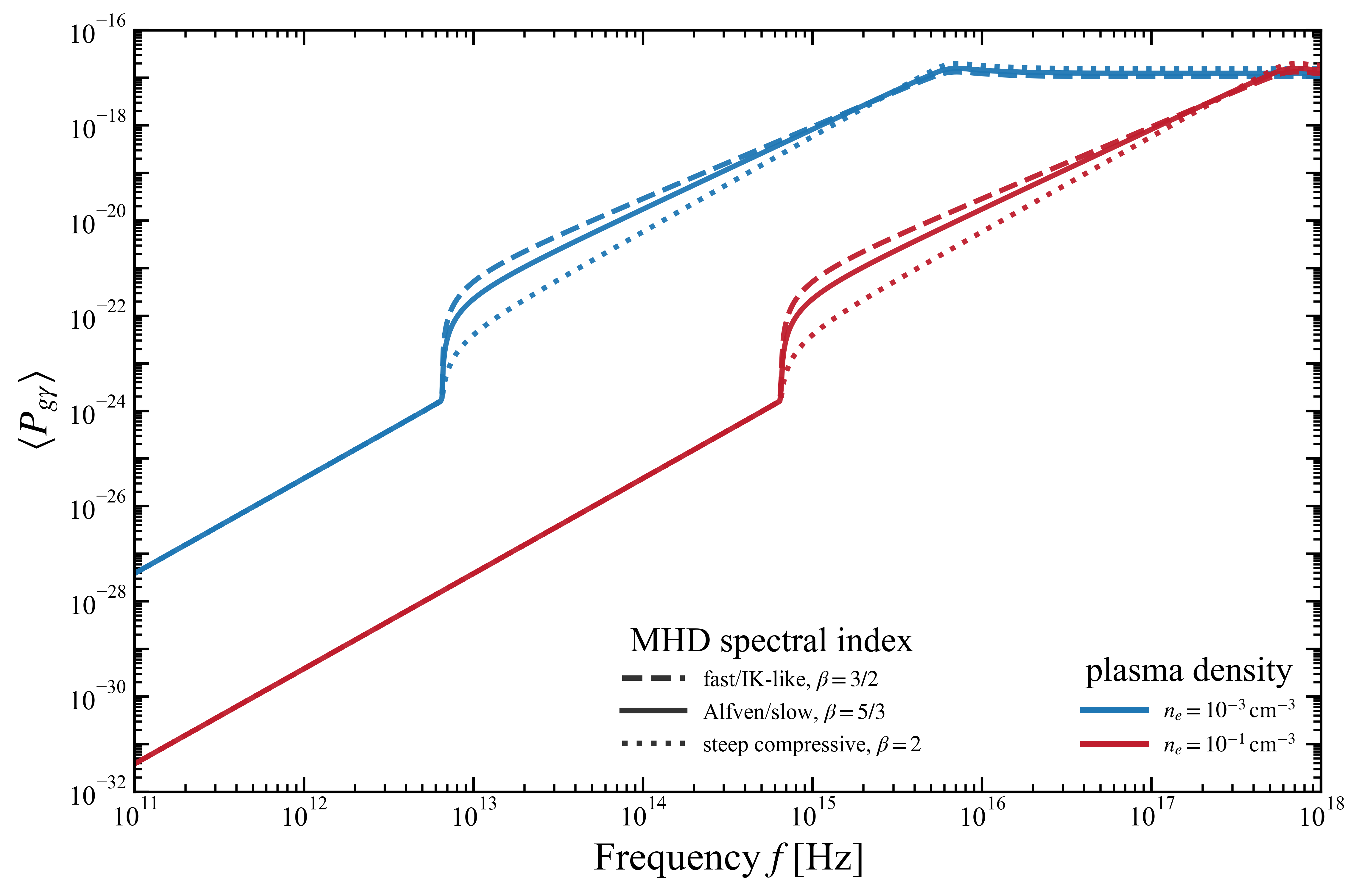}
    \caption{Finite path numerical conversion probability for MHD motivated benchmark choices of the turbulent spectral index. Colors show the two electron density benchmarks, while line styles show $\beta=3/2$, $5/3$ and $2$. The calculation uses Eq.~\eqref{eq:sup_window} with the same magnetic normalization and inertial range cutoffs as Table~\ref{tab:benchmarks}. The low frequency oscillations are finite window side lobes in the high mismatch regime. \rev{Inside the inertial range phase matching window, the slopes $f^{3/2}$, $f^{5/3}$ and $f^2$ trace the magnetic spectrum selected by the plasma mismatch.} \rev{The figure uses the inverse wave number convention $k_L^{-1}=250\,{\rm pc}$.}}
    \label{fig:mhd_beta_scan}
\end{figure}

{\bf Finite path length and the resonant limit.} The finite line of sight response follows directly from the amplitude integral. If one Fourier component of the turbulent field varies as $e^{iks}$, Eq.~\eqref{eq:sup_amplitude} contains the integral of $e^{i(k-\Delta)s}$ over the finite interval $0<s<L$. After shifting the origin to the path midpoint, this gives
\begin{equation}
    K(x)=\int_0^L\dd s\,e^{ix(s-L/2)}
    =\frac{2\sin(xL/2)}{x}.
\label{eq:sup_window_continuous}
\end{equation}
Substituting the projected correlation in Eq.~\eqref{eq:sup_projected_corr} into Eq.~\eqref{eq:sup_prob} gives the ensemble-averaged probability
\begin{align}
\langle P_{g\gamma}(\Delta)\rangle
={}&\frac{\kappa^2B_0^2}{4}
\int_0^{\kH}\dd k\,\Psi(k)\nonumber\\
&\times\left[K^2(k-\Delta)+K^2(k+\Delta)\right].
\label{eq:sup_window}
\end{align}
Here $k$ is the longitudinal wave number. We use Eq.~\eqref{eq:sup_window} for the numerical probabilities, retaining the finite propagation length. When $\Psi(k)$ varies slowly across the response width of order $L^{-1}$ around $k=\Delta$, the long path limit gives
\begin{equation}
    \lim_{L \to \infty} K^2(k-\Delta)=2\pi L\,\delta(k-\Delta).
\end{equation}
Equation~\eqref{eq:sup_window} then reduces to the resonant expression
\begin{equation}
    \langle P_{g\gamma}(\Delta)\rangle
    =
    \frac{\pi}{2}\kappa^2B_0^2L\,\Psi(\Delta) .
\label{eq:sup_res_probability}
\end{equation}
The oscillations that appear in the large-$\Delta$ tail of a direct finite path calculation are side lobes of $K(x)$, not additional physical resonances. Averaging over a frequency interval containing many oscillations recovers the phase-averaged $\Delta^{-2}$ scaling.

{\bf Asymptotic probabilities.} The plateau and inertial range expressions follow from Supplemental Eqs.~\eqref{eq:sup_psi} and \eqref{eq:sup_res_probability}, under the long path condition. The high-mismatch expression instead follows from the phase-averaged finite path response in Eq.~\eqref{eq:sup_window_continuous}.

\color{black}
For low mismatch, one finds a flat function of $\Delta$, as anticipated in the main text. In the regime $\Delta\ll\kL$, the projected spectrum is evaluated below the outer scale cutoff, so $\kmin=\kL$. Keeping the leading term in $\Delta/\kL$ and neglecting powers of $\kL/\kH$ gives
\begin{equation}
    \Psi(\Delta)\simeq
    \frac{\beta-1}{4\beta\kL}.
\end{equation}
The probability is therefore independent of $\Delta$ at leading order. This is the low mismatch plateau: although the three dimensional spectrum has a lower cutoff $\kL$, modes with total wave number $k\ge\kL$ can have longitudinal components $k_z=\Delta<\kL$. Phase matching therefore remains possible below the inertial range window.

As $\Delta$ increases through the inertial range, the projected magnetic power at the matching wave number decreases, and the ensemble-averaged conversion probability follows the corresponding spectral power law. In the regime $\kL\ll\Delta\ll\kH$, the selected longitudinal wave number is inside the inertial range, so $\kmin=\Delta$. The two terms in Eq.~\eqref{eq:sup_psi} have the same leading scaling,
\begin{align}
    \Psi(\Delta)
    &\simeq
    \frac{(\beta-1)\kL^{\beta-1}}{4}
    \left(\frac{\Delta^{-\beta}}{\beta}
    +\frac{\Delta^{-\beta}}{\beta+2}\right)
    \nonumber\\
    &=
    \frac{\beta^2-1}{2\beta(\beta+2)}
    \kL^{\beta-1}\Delta^{-\beta}.
\end{align}
Substitution into the phase-matching approximation gives the inertial range power law. Since $\Delta\propto f^{-1}$, this line becomes $\langle P_{g\gamma}\rangle\propto f^\beta$ when plotted against photon frequency. \rev{Its normalization grows linearly with $L$ as the matched longitudinal modes contribute over the propagation path.}

For yet larger $\Delta$, the now familiar $\Delta^{-2}$ scaling sets in. The phase-matching approximation has no support when $\Delta\gg\kH$, because the turbulent spectrum contains no longitudinal mode at the required wave number. The appropriate limit is instead the phase-averaged finite path tail. It is equivalent to the phase-averaged uniform field scaling with the transverse magnetic power replaced by its isotropic ensemble value, $\langle B_\perp^2\rangle=2B_0^2/3$, and with $\langle\sin^2(\Delta L/2)\rangle=1/2$. This gives
\begin{equation}
\langle P_{g\gamma}\rangle =
\begin{cases}
\displaystyle
\kappa^2B_0^2L\,\frac{\pi(\beta-1)}{8\beta\kL},
& \Delta\ll\kL,\\[1.0ex]
\displaystyle
\kappa^2B_0^2L\,
\frac{\pi(\beta^2-1)}{4\beta(\beta+2)\kL}
(\kL/\Delta)^\beta,
& \kL\ll\Delta\ll\kH,\\[1.2ex]
\displaystyle
\frac{\kappa^2B_0^2}{3\Delta^2},
& \Delta\gg\kH .
\end{cases}
\label{eq:sup_asymptotic}
\end{equation}
\color{black}
Thus the first two lines are controlled by the projected magnetic power spectrum, whereas the last line is the high mismatch dispersive tail and is independent of any length scale associated with the magnetic field configuration.

\color{black}
{\bf Photon intensity and characteristic strain.} The conversion from a photon background measurement to a GW strain bound depends on the intensity convention. Let $I_f(f,\bn)$ be the photon spectral intensity per unit frequency and per steradian. The intensity per logarithmic frequency is
\begin{equation}
    J_\gamma(f,\bn)\equiv f I_f(f,\bn).
\end{equation}
For an isotropic photon background, the corresponding energy density per logarithmic frequency is
\begin{equation}
    \frac{\dd\rho_\gamma}{\dd\ln f}
    =
    4\pi J_\gamma(f),
\label{eq:sup_photon_density}
\end{equation}
where natural units are used. If a catalogue reports $\nu I_\nu$, this quantity is precisely $J_\gamma$ after identifying $\nu=f$. If it reports $I_\nu$, one first multiplies by $f$.

The stochastic GW convention used in the main text is
\begin{equation}
    \frac{\dd\rho_{\rm GW}}{\dd\ln f}
    =
    \frac{\pi f^2}{4G}h_c^2(f).
\label{eq:sup_gw_density}
\end{equation}
For an isotropic GW background, conversion along direction $\bn$ produces a photon intensity per logarithmic frequency
\begin{equation}
    J_{\gamma}^{\rm conv}(f,\bn)
    =
    \frac{P_{g\gamma}(f,\bn)}{4\pi}
    \frac{\dd\rho_{\rm GW}}{\dd\ln f}.
\label{eq:sup_conv_intensity}
\end{equation}
The factor $4\pi$ converts an isotropic energy density into intensity per steradian. Requiring this converted component not to exceed an allowed photon intensity $J_\gamma^{\rm lim}(f,\bn)$ gives
\begin{equation}
    h_c(f)\lesssim
    \left[
    \frac{16G\,J_\gamma^{\rm lim}(f,\bn)}
    {f^2P_{g\gamma}(f,\bn)}
    \right]^{1/2}.
\label{eq:sup_hc_bound}
\end{equation}
\color{black}
In the benchmark calculation in the main text we replace $P_{g\gamma}(f,\bn)$ by the full finite path ensemble-averaged probability in Eq.~\eqref{eq:p_finite_path}, equivalently Supplemental Eq.~\eqref{eq:sup_window}. The local approximation in Eq.~\eqref{eq:p_res_full} applies only in its stated regime. This gives
\color{black}
\begin{equation}
    h_c(f)\lesssim
    \left[
    \frac{16G\,J_\gamma^{\rm lim}(f)}
    {f^2\langle P_{g\gamma}(f)\rangle}
    \right]^{1/2},
\label{eq:sup_hc_bound_avg}
\end{equation}
for the same intensity convention.

For a measurement $J_i\pm\sigma_i$ at frequency $f_i$, we write the measured intensity as the sum of a converted component, an unknown astrophysical component and noise. The converted component is
\begin{equation}
    J_{g\to\gamma}(h_c,f_i)=
    \frac{f_i^2\langle P_{g\gamma}(f_i)\rangle}{16G}\,h_c^2 .
\label{eq:sup_converted_component}
\end{equation}
Profiling over a nonnegative astrophysical intensity gives the one bin statistic
\begin{equation}
\chi_i^2(h_c)=
\begin{cases}
0,&J_{g\to\gamma}(h_c,f_i)\le J_i,\\
\left[J_{g\to\gamma}(h_c,f_i)-J_i\right]^2/\sigma_{i,+}^2,
&J_{g\to\gamma}(h_c,
f_i)>J_i .
\end{cases}
\label{eq:sup_profile_chi2}
\end{equation}
The conservative pointwise 95\% upper limit used for the measurement points is obtained from $\Delta\chi_i^2=3.84$, so that
\begin{equation}
    J_\gamma^{\rm lim}(f_i)=J_i+1.96\,\sigma_{i,+}.
\label{eq:sup_meas_limit}
\end{equation}
For a reported upper limit, we set $J_\gamma^{\rm lim}=J_i^{\rm UL}$ and keep the confidence level as reported by the source compilation. For the EBL model curve, $J_\gamma^{\rm lim}=J_{\rm EBL}(f)$, so the converted intensity is required to lie below the model intensity itself. A lower limit or galaxy count estimate does not define an upper bound on a positive converted component unless it is combined with an independent total intensity upper bound. A final data analysis must specify whether the input is $I_\nu$, $\nu I_\nu$, or a band integrated intensity, and must avoid extrapolating outside the measured frequency range.

{\color{black}
{\bf Magnetic realizations} 
The ensemble-averaged probability $\langle P_{g\gamma}\rangle$ averages over allowed magnetic configurations and does not specify the conversion probability for the particular realization of the Galactic turbulent field. Longer observations of a stochastic GW background do not sample new realizations of this fixed field. We test the ensemble-averaged conversion probability using $10{,}000$ independent realizations of a zero mean, isotropic Gaussian turbulent magnetic field with the same spectrum and normalization as the analytical calculation. We adopt $\beta=5/3$, $B_0=5\,\mu\mathrm{G}$, $k_L^{-1}=250\,\mathrm{pc}$, $k_H/k_L=10^3$, and $L=4\,\mathrm{kpc}$, with uniform electron densities $n_e=10^{-3}$ and $10^{-1}\,\mathrm{cm}^{-3}$. Since the conversion amplitudes depend linearly on the Gaussian field, we sample their joint Gaussian distribution using the covariance determined by the projected magnetic spectrum and the finite propagation length. This preserves cross frequency correlations within each realization. Figure~\ref{fig:gaussian_magnetic_realizations} compares the Monte Carlo means with the analytical predictions at frequencies spanning $10^{12}$--$10^{18}\,\mathrm{Hz}$. Figure~\ref{fig:gaussian_quantile_ratios} shows that away from the boundaries of the inertial range phase-matching window, the pointwise 5th and 95th percentiles approach $P_5\simeq0.18\langle P_{g\gamma}\rangle$ and $P_{95}\simeq2.37\langle P_{g\gamma}\rangle$ for both density benchmarks.

\begin{figure}[t]
    \centering
    \includegraphics[width=0.8\columnwidth]{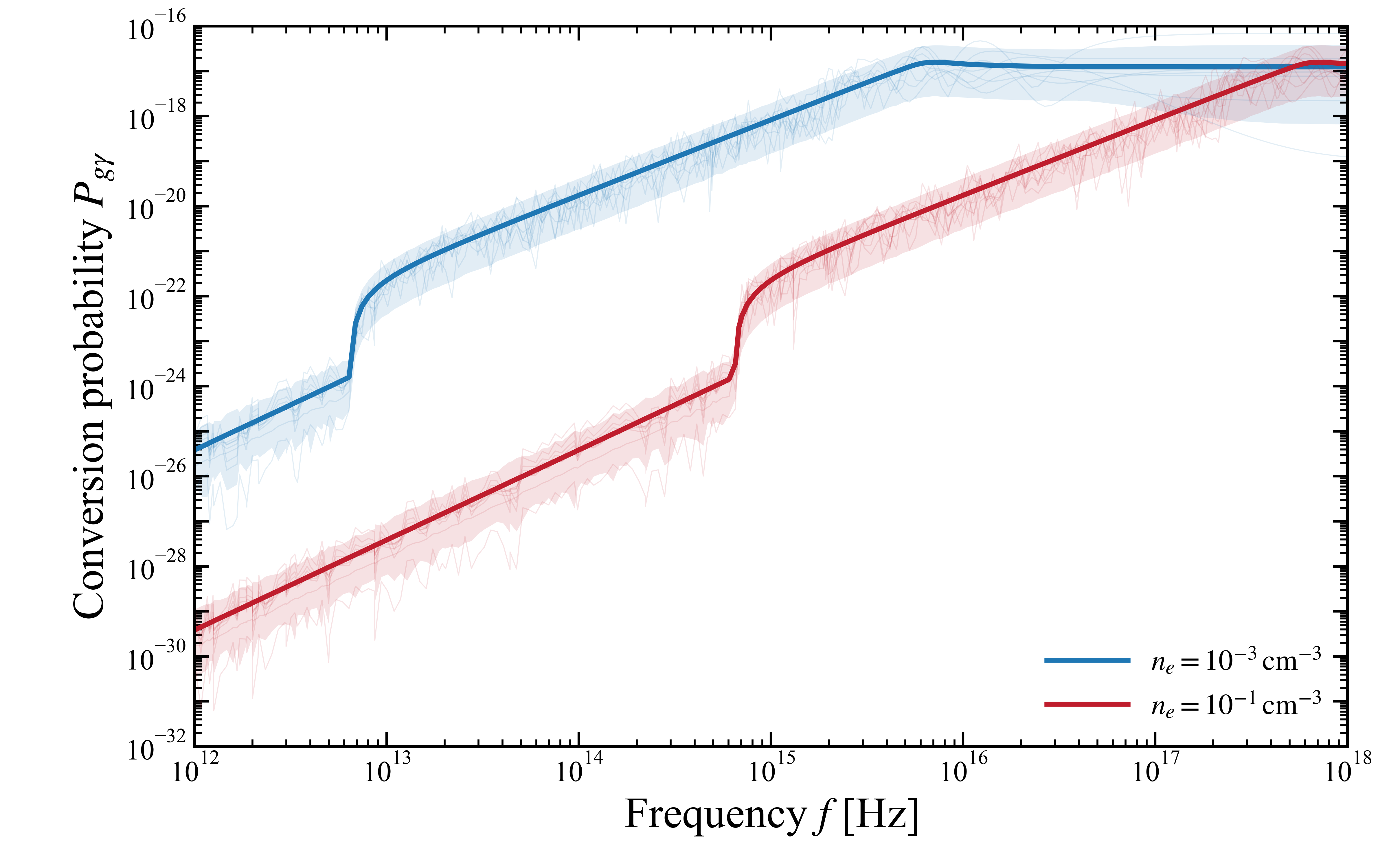}
    \caption{Graviton--photon conversion probability for $n_e=10^{-3}\,\mathrm{cm}^{-3}$ (blue) and $n_e=10^{-1}\,\mathrm{cm}^{-3}$ (red). Solid curves show the analytical ensemble averages; dashed curves show the means of $10{,}000$ Gaussian magnetic realizations. Shaded bands span the pointwise 5th--95th percentiles, and faint curves show 8 representative realizations for each density. The bands describe realization to realization scatter, not confidence intervals on the mean.}
    \label{fig:gaussian_magnetic_realizations}
\end{figure}

\begin{figure}[b]
    \centering
    \includegraphics[width=0.8\columnwidth]{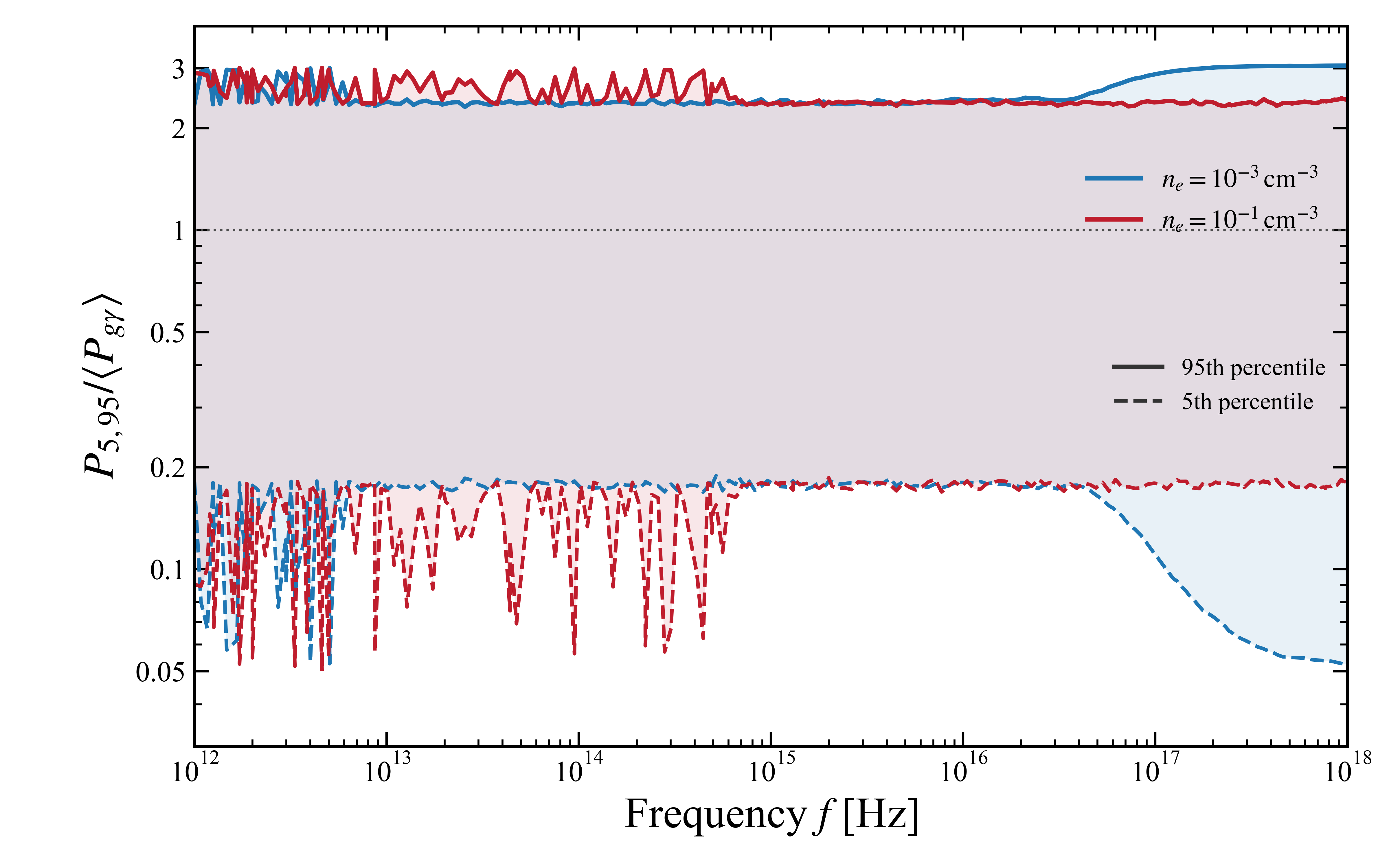}
    \caption{Pointwise realization percentiles relative to the analytical ensemble mean. Dashed and solid curves show $P_5/\langle P_{g\gamma}\rangle$ and $P_{95}/\langle P_{g\gamma}\rangle$, respectively. Shading encloses the central 90\% realization interval. The horizontal dotted line marks unity.}
    \label{fig:gaussian_quantile_ratios}
\end{figure}
}

\end{document}